\documentclass[traditabstract]{aa} % for the abstract without structuration 
\usepackage{graphicx}
\usepackage{txfonts}
\begin{document}
%Mira properties -- specify better (radius, system, pulsation)

  \title{The spectroscopic  evolution of the symbiotic-like recurrent nova V407 Cygni during its 2010 outburst}
   \subtitle{II. The circumstellar environment and the aftermath}

   \author{S. N. Shore\inst{1,2}, G. M. Wahlgren\inst{3,4}, T. Augusteijn\inst{5}, T. Liimets\inst{5,6}, P. Koubsky\inst{7}, M. \v{S}lechta\inst{7}, V. Votruba\inst{7}    }
        \institute{
   Dipartimento di Fisica ``Enrico Fermi'', Universit\`a di Pisa, largo B. Pontecorvo 3, I-56127 Pisa, Italy\\ \email{shore@df.unipi.it}
   \and
   INFN - Sezione di Pisa
   \and  
   The Catholic University of America, Dept. of Physics, 620 Michigan Ave NE, Washington DC, 20064, USA\\   \email{glenn.m.wahlgren@nasa.gov}
      \and
NASA-GSFC, Code 667, Greenbelt, MD, 20771, USA
\and
Nordic Optical Telescope, Apartado 474, E-38700 Santa Cruz de La Palma, Santa Cruz de Tenerife,
Spain\\ \email{tau@not.iac.es,tiina@not.iac.es}
\and
Tartu Observatory,  T\~oravere, 61602, Estonia
 \and
  Astronomical Institute, Academy of Sciences of the Czech Republic,  CZ-251 65   Ond\v{r}ejov, Czech Republic\\ \email{koubsky@sunstel.asu.cas.cz}      
  }
     
              \date{submitted: -- ;  accepted ---}
%DRAFT Revision: 18/11/11; 16/8/11; 20/5/11

%\abstract{}{}{}{}{} 
% 5 {} token are mandatory
 \abstract
{The nova outburst of V407 Cyg in 2010 Mar. 10 was the first observed for this star but its close resemblance to the well known symbiotic-like recurrent nova RS Oph suggests that it is also a member of this rare type of Galactic novae.  The nova was the first detected at $\gamma$-ray energies and is the first known nova explosion for this system.   The extensive multiwavelength coverage of this outburst makes it an ideal comparison with the few other outbursts known for similar systems.   
   We extend our previous analysis  of the Mira and the expanding shock from the explosion to detail the time development of the photoionized Mira wind, circumstellar medium, and shocked circumstellar environment to derive their physical parameters and how they relate to large scale structure of the environment, extending the previous coverage to more than 500 days after outburst.  We use  optical spectra obtained at high resolution with the Nordic Optical Telescope (NOT) (R$\approx$45000 to 65000) and medium resolution Ond\v{r}ejov  Observatory (R$\approx$12000) data and compare the line variations  with publicly available archival measurements at 30 GHz OVNR and at X-rays with Swift during the first four months of the outburst, through the end of the epoch of strong XR emission.   We use nebular diagnostics and high resolution profile variations to derive the densities and locations of the extended emission.  We find that the higher the ionization and/or the higher the excitation energy, the more closely the profiles resemble the He II/Ca V-type high velocity shock profile discussed in Paper I.  This also  accounts for the comparative development of the [N II] and [O III] isoelectronic transitions: the [O III] 4363\AA\ profile does not show the low velocity peaks while the excited [N II] 5754\AA\ does.  If nitrogen is mainly N$^{+3}$ or higher in the shock, the upper state of the [N II] nebular lines will contribute but if the oxygen is O$^{+2}$ then this line is formed by recombination, masking the nebular contributor, and the lower states are collisionally quenched but emit from the low density surroundings.    Absorption lines of Fe-peak ions formed in the Mira wind were visible as P Cyg profiles at low velocity before Day 69, around the time of the X-ray peak and we identified many absorption transitions without accompanying emission for metal lines.   The H Balmer lines showed strong P Cyg absorption troughs that weakened during the 2010 observing period, through Day 128.   The Fe-peak  line profiles and flux variations were different for permitted and forbidden transitions: the E1 transitions were not visible after Day 128 but had shown a narrow peak superimposed on an extended (200 km s$^{-1}$) blue wing,  while the M1 and E2 transitions persisted to Day 529, the last observation, and showed extended redshifted wings up of the same velocity.    We distinguish the components from the shock, the photoionized environment, and the chromosphere and inner Mira wind using spectra taken more than one year after outburst.  The multiple shells and radiative excitation phenomenology are similar to those recently cited for GRBs and SNIa.}


          \thanks{Based on observations made with the Nordic Optical Telescope, operated on the island of La Palma jointly by Denmark, Finland, Iceland, Norway, and Sweden, in the Spanish Observatorio del Roque de los Muchachos of the Instituto de Astrofisica de Canarias. }
            \titlerunning{The 2010 outburst of the recurrent nova V407 Cyg. II}   \authorrunning{S. N. Shore et al.}
   \maketitle

\section{Introduction}

The 2010 outburst of V407 Cyg, a previously classified D-type
symbiotic star (Munari, Margoni, \& Stagni 1990) whose historical
variability had been classified as ``nova-like'' by Hoffmeister (1949)
and Meinunger (1966), developed photometrically and spectroscopically almost identically to the most recent
eruptions of the symbiotic-like recurrent nova (SyRN) RS Oph (1985, 2006).  This event 
moved the system into that rare group of explosive cataclysmic
variables.   In Shore et al. (2011, hereafter  paper I)  we
presented the first results of the outburst based on multiwavelength
photometry and spectroscopy from the Nordic Optical Telescope,
Ond\v{r}ejov Observatory, {\it Swift}, and {\it Fermi/LAT} concentrating on the Mira
properties and the time development of the shock propagating through
the red giant (RG) wind.  In this second paper, we focus on the evolution of the
circumstellar environment  that was photoionized and illuminated by
the emission from the white dwarf (WD) and the shock.  We extend the previous results
to include additional proxies for the high energy excitation mechanisms.

%\justification 
%__________________________________________________________________

\section{Observational data}

Our principal optical spectroscopic  data set, the same as Paper I, consists of
spectra taken between 2010 Mar 24 and 2010 Jul 16 with the
Ond\v{r}ejov Observatory Zeiss 2.0 m telescope coud\'e spectrograph
and with the 2.6 m Nordic Optical Telescope (NOT) fiber-optic echelle
spectrograph (FIES, program P40-423).  The Ond\v{r}ejov spectra taken with the SITe005 800x2000 chip were mainly obtained in the vicinity of
H$\alpha$,  supplemented by spectra at both bluer (4000-5000\AA) and
redder (8000-9000\AA) wavelengths with a dispersion of 0.24\AA\ px$^{-1}$ and a
spectral coverage of $\approx$500\AA\ with  exposure times ranging
from 60 s to 6700 s. The FIES spectra were obtained with a dispersion
of 0.023\AA\ px$^{-1}$ in high-resolution mode covering  from 3635\AA\ to
7364\AA\ and 0.035\AA\ px$^{-1}$ in medium-resolution mode covering
3680\AA\ to 7300\AA.  Exposures ranged from 100 s to 3000 s to correct for
overexposure at H$\alpha$.   Some adjacent orders display ``phantom''
H$\alpha$ profiles and we have avoided those spectral intervals in
this analysis.  Absolute fluxes for the NOT spectra were obtained
using the flux standard star BD +28$^o$4211 observed on 2010 Jun 3 (JD
55370) and Jun 23 (JD 55378) at high resolution, the rest of the
sequence was not absolutely calibrated.  All NOT spectra were reduced
using IRAF, FIESTools, and IDL.\footnote{ IRAF is distributed by the
National Optical Astronomy Observatories, which are operated by the
Association of Universities for Research in Astronomy, Inc., under
cooperative agreement with the US National Science Foundation.}  On
several dates, contemporaneous spectra between the two observatories
allow us to correlate and cross-calibrate the data.   The wavelengths were checked against the interstellar Ca II H and K and Na I D1 and D2 lines.  
The journal of our observations was presented in Paper I (Tables 1a and 1b) and
will not be repeated here.  As in Paper I, when referring to {\it date
after outburst} we adopt as the date of X-ray onset 2010 Mar 10.813 = JD 2455266.313 from the {\it
Swift} XR analysis (see also CBET \#2199).

A late-time high resolution FIES spectrum, discussed in the last section of this paper, was obtained on 2011 Aug. 21.2 (JD 55784.7) with an exposure time of 4400 sec.  The spectrum covers the region 3700 - 7300\AA\ but the signal-to-noise ratio is very low below 4700\AA\ and we concentrate only on the longer wavelength region.   This spectrum was fluxed by comparing it with a calibrated spectrum at lower resolution obtained at  Ond\v{r}ejov on 2011 Apr. 23 (JD 55674.7) along with an exposure of the flux standard BD +28$^o$4211 in the wavelength interval 5736 - 6765\AA.  The change in the absolute brightness of the system between these two dates was likely small.  The NOT spectrum was convolved to the resolution of the 23 Apr. spectrum and scaled to match the intensities of the profiles of the strongest lines (e.g. [O I] 6300\AA, He I 5875\AA).  The resulting systematic flux uncertainty appears to be less than 30\%.

\section{The first 150 days of outburst: 2010}

We first present several results that extend the analysis of the shock
propagation and then pass to the derivation of the physical properties
of the photoionized circumstellar medium.  The two are linked by the
radiation from the ejecta propagating through the Mira wind.
In Paper I we identified three separate types of emission line
profiles, originating from different mechanisms and locations.  The
narrowest component, with a FWHM of about 10 km s$^{-1}$, we
identified with the chromosphere.  The [O I] lines showed this most
clearly in the first NOT spectrum but these also developed broad
wings; in contrast, the [Ca II] 71291, 7323\AA\ profiles remained
stationary and extremely narrow throughout the observed period.  The
second component is the emission and absorption from the RG wind and
photoionized circumstellar emission that displayed relatively narrow
components with FWHM$\approx$40 km s$^{-1}$ due to fluorescence and
emission from the wind in the form of either pure emission, pure
absorption, or narrow P Cyg profiles.  These were observed on the
Balmer profiles and are due to heavy metal ions.  The last
component, which developed differently depending on the species, was
from the shock as it propagated through the RG wind.  This was
the main result of Paper I, that the shock -- as in RS Oph  -- powered an expanding ionized region that eventually engulfed  the
entire peripheral region of the Mira wind and also accelerated the material to several hundred km s$^{-1}$.  

\subsection{Shock interaction with the wind: the high velocity component on the emission lines}

No He I singlet transition was detected {\it except} 6678\AA.   The
triplets detected were 4471, 4713, 4921, 5875, and 7065\AA; for those
at shorter wavelength the signal to noise ratio was too low for
detection.   The profiles developed along the same lines as He II
4686\AA, with an increasingly evident redward extension to +350 km
s$^{-1}$ and a blueward wing that decreased from -500 km s$^{-1}$ to
-200 km s$^{-1}$ from JD 55288 to JD 55397.  The strong He I triplets,
4471\AA\ and 5875\AA, also showed a narrower component that coincided 
in radial velocity with, and is similar in profile to,  the narrow peaks on the
nebular [N II] and [O III] lines.  We will discuss this further in
sec. 3.4.1 and 4, below.  We show in Fig. 1 the comparison of the profiles of the He I 5875\AA\
triplet with He II 4686\AA\ and [N II] 6548\AA.  After JD
55350 these profiles remained almost constant in form.  The two He I
peaks corresponded to the same displacement, -23$\pm$2 km s$^{-1}$,
relative to the rest wavelength of the Mira.  The redshifted
component remained relatively weaker in He II than the He I or the 
nebular lines with a similar intensity ratio of the blue to red
component (once the underlying shock profile was removed).   The He II 5411\AA\ line was present but blended with [Fe II] 5412.65\AA.  It produced the blue wing on the forbidden line, extending to -200 km s$^{-1}$, that  was detected in the last two NOT spectra from 2010 and the 2011 Aug. 21 spectrum.  We discuss this further in sec. 4.

Comparing the [Ca V] 5309\AA\ line to He II 4686\AA\ provides a means for disentangling the environmental and shock contributions to the profile.  This is made difficult by the blend of [Ca V] with the wing of Fe II 5316.6\AA,  but there was sufficient separation near the line center that the peak and blue wings can be compared.  From this, it appears that the He II 4686\AA\ line also showed a contribution from the circumstellar gas coincident in radial velocity with those observed on the nebular lines (see below) at low velocity.  

    \begin{figure}
   \centering
   \includegraphics[width=7cm]{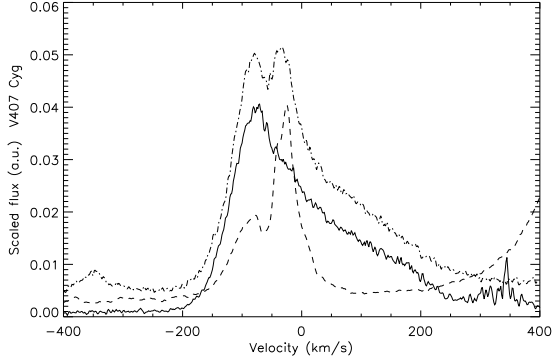}
   \caption{Comparison of [N II] 6548\AA\ (dash), He II 4686\AA\ (solid), and He I 5875\AA\ (dot-dash) profiles on JD 55370, near the end of the NOT sequence.  The resolved peaks are due to the ionized region in the Mira wind and the extended wind is from the shock.  The profiles have been scaled in intensity for comparison  (see discussion).}
    \end{figure}

%% return to the He I triplet discussion in sec. on nebular diagnostics

The [Fe X] 6347\AA\ line was detected in the early outburst of several SyRNe, especially V745 Sco 1989 (e.g. Williams et al. 2003).  As we show in  Fig. 2, the line was also detected in V407 Cyg.  It followed precisely the XR light curve from {\it Swift} that we discussed in Paper I.  The intensity maxima were essentially simultaneous, the line weakened following the X-rays, and then after about Day 60 remained at nearly constant equivalent width (and flux, at that stage the additional continuum had all but disappeared, see Paper I).     The line profile remained asymmetric, similar to He II and [Ca V], but the redward extension decreased in maximum velocity from -300 km s$^{-1}$ to about -200 km s$^{-1}$ relative to the rest velocity.   The effect remains even after the continuum emission, that is presumed to be from the shock and that masks the Mira photosphere, was removed.  This is the correction factor (CF) listed in Table 2, determined using the photospheric Li I resonance line (see Paper I).   In contrast, as we previously discussed,  [Fe VII]  showed continued profile evolution and a broad, shell-like structure.   The [Fe X] line was absent in the late epoch spectrum obtained on 2011 Apr. 23 (JD 55674) at Ond\v{r}ejov with an upper limit to the equivalent width of  5.4$\pm$0.2\AA.  There were no contemporaneous {\it Swift} observations during this last observation.    {\it The strong correlation between the X-ray emission and [Fe X] argues, however, that the source had turned off in the XR band by that time and highlights the utility of this forbidden line as a shock diagnostic in other symbiotic systems even in the absence of observable X-rays}.   The highest confirmed ionization state was Fe$^{+9}$ .   For completeness, we also list ground state lines for the high ionic species that were {\it not} detected during the NOT sequence, most of which were reported in the post-outburst spectrum of RS Oph by Wallerstein \& Garnavich (1986) with lower resolution data, 1 to 4\AA\ at a time corresponding to our spectrum on JD 55338, about Day 70 of the outburst: [Ar IV] 4711\AA\footnote{The identification of [Ar IV] 4711\AA\ given in Paper I was wrong.   The line is He I 4713.14\AA, a triplet transition, that showed a [Ca V]-like profile after JD 55334 and was not detected before JD 55288.}; Mn V] 6024, 6083\AA; [Mn XII] 6660\AA; [Mn XIII] 6536\AA; [Cr IV] 7051, 7110\AA; Fe VI] 5097, 5277, 5335\AA; [Fe XI] 6985\AA; [Fe XIV] 5302; [Fe XIII] 5386\AA; [Ni XII] 4230\AA; [Ni XII] 5115.8\AA.\footnote{For the lower ionization lines on that date after outburst in RS Oph, Wallerstein \& Garnavich report radial velocities that are all low velocity, -37$\pm$4 km s$^{-1}$, the coronal lines were blueshifted from this by 160$\pm$5 km s$^{-1}$.  They did not report emission from Sc II but did detect Si II, Ti II, and Fe II in the late-time spectra.}  
\begin{center}
{\bf Table 1. [Fe X] 6374\AA\ equivalent widths}\\
\begin{tabular}{ccc}
\hline
JD240000+ & CF & EW ([Fe X]) (\AA) \\
\hline
55279 &       3.0 &       16.4\\
      55288&       3.0 &       24.3\\
      55294 &       2.7 &       51.6\\
      55308 &       2.1 &       60.6\\
      55314 &       1.8 &       43.4\\
      55315 &       1.7 &       57.6\\
      55334 &      0.18 &       9.3\\
      55350 &     0.09 &       7.6\\
      55370 &     0.01 &       6.8\\
\hline
\end{tabular}
\end{center}

The [Fe III] 4754.69\AA\ line  was isolated and easily measured during the outburst.  While it resembled the shock-dominated profiles, beginning with a blueward extension to -100 km s$^{-1}$ and then changing the sense of asymmetry to develop a redward extension to 350 km s$^{-1}$  relative to the Mira, the profile was more symmetric by JD 55370.  The equivalent width peaked after that of [Fe X] 6374\AA\, around JD 55340 (about one month after [Fe X] and the X-rays), indicating that recombination in the post-shocked gas was responsible for the line.  It never displayed the high velocities of the lines from H, He I, He II, or highly ionized metallic species.

     \begin{figure}
   \centering
   \includegraphics[width=8cm]{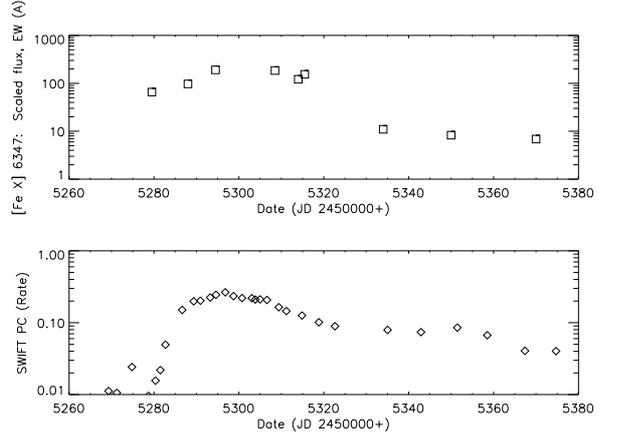}
   \caption{ Top: Equivalent width (\AA) variations of the [Fe X] 6374\AA\ shock line corrected for the contribution of the excess emission from the shock.    bottom: {\it Swift PC} count rate during the same sequence (see text for details). }
    \end{figure}
    
\subsection{Balmer line wind components}

The H$\alpha$ and H$\beta$ profiles show a low velocity absorption that remained strong throughout the period covered by these observations but weakening progressively and narrowing (see Table 1 and Fig. 3).  For the H$\gamma$ and H$\delta$ profiles, however, after JD 55370 emission began to fill in the absorption and for H$\delta$ the absorption disappeared with an almost mirror image emission line replacing it at the same velocity.  The rest of the line profile developed similarly for all four Balmer lines.

\begin{center}
{\bf Table 2. Balmer line velocities (km s$^{-1}$)}\\
\begin{tabular}{c|ccc|ccc}
\hline
 &  & v$_{rad,max}$ &  & & v$_{rad,min}$  & \\
 Date (JD) & H$\alpha$ & H$\beta$ & H$\gamma$ & H$\alpha$ & H$\beta$ & H$\gamma$ \\ 
\hline
55288 &  -125 & -107& -86 &   -67 & -58 & -51 \\
55370 & -103 & -84 & -77  &    -71 & -65 & -61 \\
55378 & -100 & -82 & -79 &   -69 & -66 & -61 \\
\hline
\end{tabular}
\end{center}

The absorption showed a maximum velocity of about -65 km s$^{-1}$ with respect to the star and decreased along the series following the Balmer progression in oscillator strength.   In the last NOT spectra of the 2010 sequence the H$\delta$ line displayed emission at the same radial velocity as the earlier P Cyg absorption, at a velocity similar to  that observed on the forbidden lines; this was probably caused by recombination in the Mira wind.  We suggest that the disappearance of the absorption component on the P Cyg profile was due to the increasing ionization of the wind and the transparency of the Lyman series, hence the passage to an optically thin ionized region that had reached the boundaries of the circumstellar environment.

    \begin{figure}
   \centering
   \includegraphics[width=8cm]{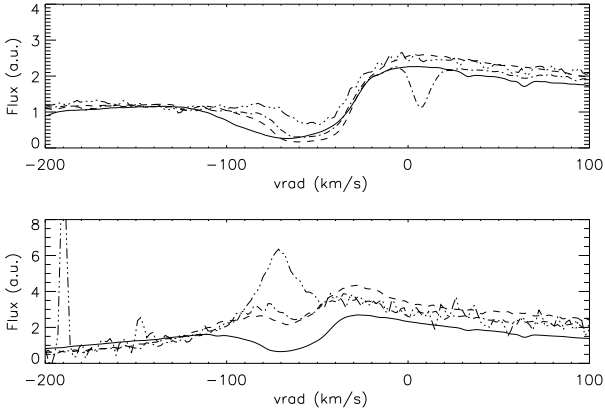}
   \caption{Comparison of the Balmer line  profiles for the V407 Cyg NOT spectra for JD 55288 (top) and JD 55370 (bottom).  The four Balmer lines are shown as: solid, H$\alpha$; dash, H$\beta$; dot-dash: H$\gamma$; dot-dot-dash, H$\delta$.(dash)}
    \end{figure}

\subsection{Low ionization metallic absorption lines from the wind}
  
Additional absorption lines on H$\beta$ and H$\gamma$ in the first spectra were narrow and not seen after JD 55314, see Fig. 4.  They all decreased in strength between the first two spectra.  The line on H$\gamma$, which had a corresponding feature on H$\beta$  at +131 km s$^{-1}$, displayed emission on the second NOT spectrum.  Thereafter, the component also vanished from the H$\beta$ line.   The decrease in the absorption component on all Balmer lines was systematic in time.  The H$\delta$ profile changed to emission in the last spectra from 2010.  The minimum velocity corresponded roughly to the Na I absorption component at -55 km s$^{-1}$ but is always blueward. Na I never shows a P Cyg profile; however, the singly-ionized Fe-peak optically thick transitions did (e.g. the RMT 42 spectrum shown in Paper I and Fig. 21, below).  We suggest that this behavior is similar to the disappearance of the overlying Fe-curtain lines on C IV 1548, 1550\AA\ during the 1985 outburst of RS Oph that Shore et al. (1996) attributed to the advancing ionization front in the RG wind.  There were no ultraviolet spectra obtained during the 2010 V407 Cyg outburst.  However, the persistence of some of the emission lines, indeed their strengthening during the development of the shock emission, argues that they arise in the chromosphere (formed from chromospheric thermal emission and fluorescence with the UV continuum) and from that portion of the Mira that was never exposed to ionizing radiation.  The narrow absorption features originated in the wind plus chromospheric absorption in the line of sight toward the shock.  This is supported by the radial velocities measured for the individual features.  The strong line [Ca II] 7323\AA, for instance, served to determine the velocity of the chromosphere and inner wind.  It was well fitted by a {\it single gaussian} profile with centroid velocity  -53.0$\pm$0.01 km s$^{-1}$ and FWHM = 10.8 km s$^{-1}$.    The mean radial velocity for the lines listed in Table 3 was -56.0$\pm$2.9 km s$^{-1}$ (-55.4$\pm$1.9 km s$^{-1}$ excluding the Fe II lines).  The uncertainty on individual measurements was $\pm$ 1 km s$^{-1}$.   The laboratory wavelengths are from the NIST atomic spectroscopy database. 

\begin{center}
{\bf Table 3. Metallic absorption line radial velocities (km s$^{-1}$)}\\
\begin{tabular}{ccc|ccc}
\hline
Ion & $\lambda_{\rm lab}$ (air, \AA) & v$_{\rm rad}$ & Ion & $\lambda_{\rm lab}$ (air, \AA) & v$_{\rm rad}$ \\
\hline
Mg I & 5167.32 & -54   & Cr I & 5208.42 &  -55 \\
Mg I & 5172.68 & -53 &  Cr II & 4856.19 & -53 \\
Mg I & 5183.60 & -54 & Cr II & 4864.33 & -54 \\
\hline
Ca I & 4226.73 & -55  & Fe II & 4177.69 & -66 \\
Ca I & 6122.22 & -54  & Fe II  & 4233.17 &  -61 \\
Ca I & 6162.17 & -54  &  & & \\
\hline
Sc II & 4246.82 & -58  & Sr II & 4077.71 & -55 \\
Sc II & 4325.00 & -55  & Sr II  & 4215.52 & -54  \\
Sc II & 5031.02 & -54  &Y II  & 4854.87  & -54 \\
Sc II & 5526.79 & -55  & Y II & 4883.69  & -54 \\
Sc II & 5657.90 & -57  & Y II & 5087.44 & -54 \\ 
Sc II & 5667.15 & -55  & Y II & 5200.41 & -54 \\
Sc II & 5669.04 & -56  & Y II & 5205.72 & -54 \\
Sc II & 5684.20 & -55  & Y II & 5662.92 (?) & -54 \\
Sc II & 6604.60 & -53  & & & \\
\hline
Ti II & 4443.79 & -60   & Ba II & 4554.03 & -55 \\
Ti II & 4468.51 & -59  & Ba II & 5853.69 & -54 \\
Ti II & 4395.03 & -59  & Ba II & 6141.71 & -54 \\
Ti II & 4344.29 & -54  & Ba II & 6496.87(:) & -55 \\
Ti II & 4337.92 & -57  &  & & \\
Ti II & 4301.91 & -57  & & & \\
Ti II & 4855.91 & -53 & & & \\
Ti II & 4865.61 & -54  & & & \\
\hline
\end{tabular}
\end{center}

    \begin{figure}
   \centering
   \includegraphics[width=8cm]{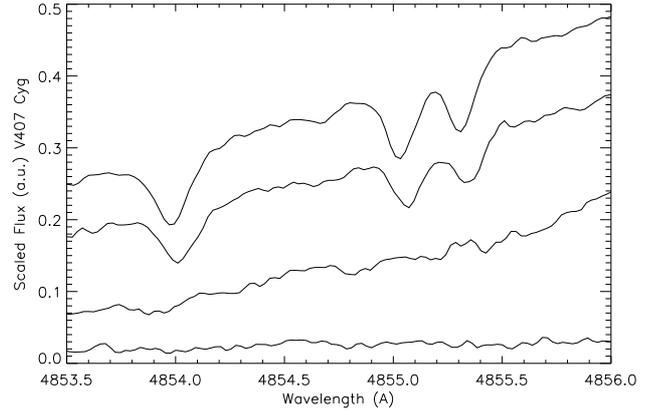}
   \caption{Development of the absorption lines in the blue wing of H$\beta$ emission line profile, Y II 4854.87, Ti II 4855.91, and Cr II 4856.86\AA, during the early stages of the outburst.  From top to bottom, the sequence is JD  55286, 55288, 55314, and 55334 (see text and Table 3 for details). }
    \end{figure}
    
% the Ba II 6497 is contaminated on the red side by terrest. water

The Na I D lines (5889.95, 5893.92 \AA) connect the ground state to the 2p$^6$3p $^2$P$^o$ levels that are fed from a common upper singlet state 2p$^6$5s $^2$S at 33201 cm$^{-1}$.  Two lines of the upper state, 6154.22, 6160.75 \AA, that have atomic transition probability $A$-values\footnote{The transition probabilities are from the NIST Atomic Spectra Database, URL: http://www.nist.gov/pml/data/asd.cfm}  of 2.5$\times 10^6$ s$^{-1}$ and 5.0$\times 10^6$ s$^{-1}$, respectively, were both detected throughout the NOT sequence with the optically thin ratio.  These are shown in Fig. 5.   Lines of the $^2$P$^o$-$^2$D multiplet, at 5682\AA\ and 5688\AA\  transition array that also feed the upper levels of the Na I D lines were not detected.  The higher excitation lines never displayed the broad emission seen on the D lines; they both remained narrow and at the system radial velocity throughout the NOT sequence.  

The Na I D lines present a problem of interpretation.  Only Mg I 5167.32\AA\ showed this double structure at the Mira velocity or had distinct wind and chromospheric/photospheric components.  If, instead of two absorption components, the feature at -54 km s$^{-1}$ was  chromospheric emission, its velocity would agree with both the metallic ion absorption lines and the [Ca II] 7291, 7323\AA\ lines.  We therefore tenatively identify the absorption with the wind and ``peak'' as persistent chromospheric emission.  The Na I absorption line cannot be photospheric in origin since it would show the same equivalent width variations (strengthening over time) as the Li resonance line.  If it is due to the wind, its strength indicates a large column density of the Mira wind.   The Mg I triplet 5167, 5172, and 5183\AA\ were among the most  persistent absorption features; the 5167.32\AA\ line was still detectable  through JD 55314.   As we noted, this transition showed a similar profile to the Na I D lines, with a weak central reversal, but with a velocity for the redward  feature of -43 km s$^{-1}$.  There was a weak excess absorption on Na I D1 and D2 at the same velocity.  In Table 3 we give the radial velocity of the component that matched the other two Mg I absorption lines, neither of which showed the double structure (see sec. 3.4.2).  

\begin{figure}
   \centering
   \includegraphics[width=9cm]{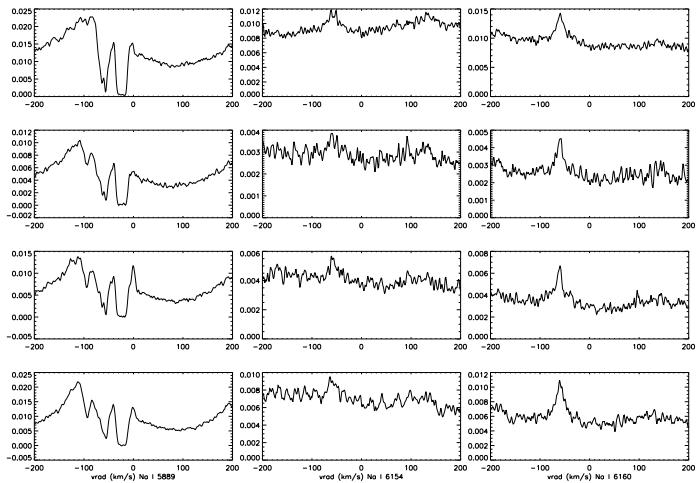}
   \caption{Comparative development of Na I  line profiles: Left: 5889.95\AA; middle: 6154.222\AA; right: 6160.75\AA, for JD 55314, 55334, 55370, and 55378.  The emission at around 0 km s$^{-1}$ and the additional absorption at around -100 km s$^{-1}$ were not observed on the higher excitation lines.}
              \label{outburst-spectra}%
    \end{figure}

We show in Fig. 6 an example of the narrow lines in the 5600-5700\AA\ region, that were in absorption from the first spectrum obtained by C. Buil\footnote{The spectra, with a resolution of 0.1\AA, were obtained from the URL: www.astrosurf.com/aras/V407Cyg/v407cyg.htm.  The  echelle data on Mar. 14,  Mar. 16, and Mar. 23  were the combination of 10, 11, and seven 600 sec sub-exposures, respectively.}  from Day 2 through  Day 30 after outburst and then changed into emission lines.  The equivalent width measurements, shown in Fig. 7 have not been corrected for the continuum since for these lines we do not yet have a firm identification of the site of formation.  The two strong absorption lines at 5667\AA\ and 5669\AA\ are Sc II arising from a common lower level of 3p$^6$3d$^2$$^3$P$_1$ at 12101.50 cm$^{-1}$ that is connected to the ground state by an E2 transition at 8261.17\AA\ that  was outside of our spectral range.  The 5667.16\AA\ line has a common upper level with Sc II 5684.21\AA\ that also goes over into emission at the same time.  The upper states of all three lines are fed from the ground state by transitions at 3361.26\AA\ and 3361.93\AA.  The emission suggests that the line formation zone had shifted by that stage to the outer wind.  

 \begin{figure}
   \centering
   \includegraphics[width=9cm]{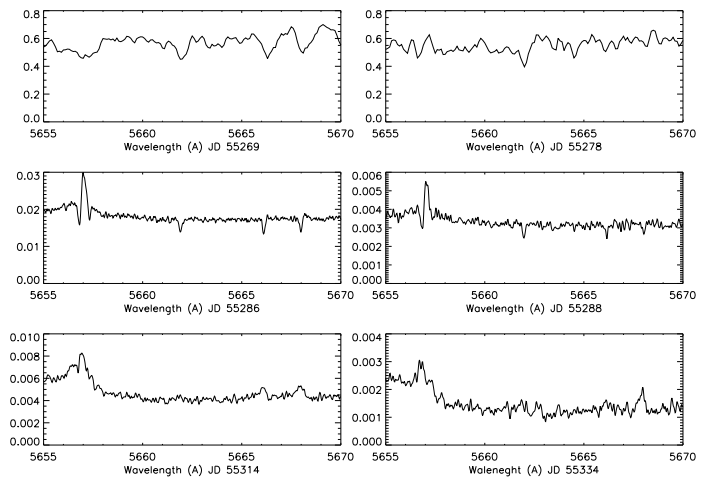}
   \caption{Example of the transformation from absorption to emission for metallic ion narrow absorption lines.  The top two spectra, echelle data from Buil, have lower resolution than the subsequent NOT spectra.  Note the change of the Sc II doublet, the 5684\AA\ line is not shown (see text). }
              \label{outburst-spectra}%
    \end{figure}

 \begin{figure}
   \centering
   \includegraphics[width=7cm]{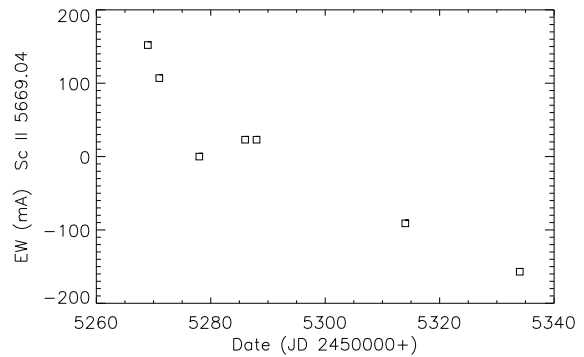}
   \caption{Equivalent width variations for Sc II 5669\AA.  The uncertainty is $\pm$10 m\AA\ for the first three spectra (the Buil data), and $\pm$ 2 m\AA\ for the NOT data.  Negative values indicate emission.  None of the measurements were corrected for the excess continuum contribution.  The Buil spectra were normalized.}
              \label{outburst-spectra}%
    \end{figure}

\subsection{Nebular emission lines from low ionization species}
\subsubsection{He, C, N, O}

No optical emission lines of any ionization state of carbon were detected during the entire outburst.

The oxygen lines presented a rare view of the detailed structure of the inner wind of the RG.   Figure 8 shows the variations in emission line equivalent width for the three neutral lines based only on the NOT spectra and Figure 9 shows a comparison for the narrow components between [O I] 6300\AA, [O III] 5007\AA, [N II] 6548\AA, and Na I D.\footnote{Figure 8 of Paper I was based on lower resolution Ond\v{r}ejov spectra that cover a shorter interval but includes the 6446\AA\ line.}  The wings of the profiles agree beyond about 100 km s$^{-1}$ from line center.  While a first interpretation of the [N II] and [O III] profiles would be either separate components with $\Delta v$ $\approx$ 40 km s$^{-1}$, the [O I] shows that the interval around the system velocity is actually a neutral zone in the wind.  There is an almost perfect coincidence of the peaks in O I and dips in the ionized lines.  The [O III] and [N II] further differ in the components at -21 and -40 km s$^{-1}$, the ratio of which is larger for the [O III] N1 and N2 lines.  
        
The 2p$^4$ $^1$D - 2p$^4$ $^1$S O I 5577.34 \AA\ line was systematically shifted from the two ground state transitions by -15$\pm$1 km s$^{-1}$.  It always showed an extended wing that increased in strength along with the 6300 \AA\ and 6363 \AA\ lines.  The initial narrow peak of the [O I] transition did not appear on this line (FWHM([O I] )= 12 km s$^{-1}$, with low level, extended wings that are similar to the excitated state transition).    The dimple in the peak of the 6300\AA\ line is due to atmospheric water vapor as are the weak absorption features on the extended line wings between -500 and +500 km s$^{-1}$ based on a comparison with a high resolution NOT spectrum obtained for calibration of BD +28$^o$4211.  The clue to its identification is also provided by the 5577\AA\ central peak that showed no corresponding feature.  The development of O I 5577.34\AA\  was   particularly interesting.  The narrow core at -54 km s$^{-1}$ (FWHM$\approx$10 km s$^{-1}$) decreased in relative strength and started to broaden before  JD 55314.  The line already displayed low level wings extending to $\pm$200 km s$^{-1}$ whose relative contribution increased until  JD 55334, after which they remain stable extending from -200 km s$^{-1}$ to 300 km s$^{-1}$.  A core component was present on the late spectra with a velocity of -63$\pm$2 km s$^{-1}$, indicating that it was from the neutral wind and not from the chromosphere.  Its radial velocity agreed with the emission components on H$\delta$.

    \begin{figure}
   \centering
   \includegraphics[width=8cm]{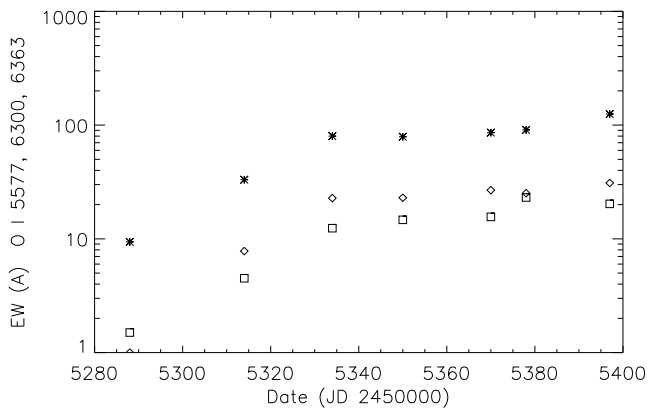}
   \caption{Equivalent width variation based on the NOT high resolution spectra of the [O I] emission lines: 5577\AA\ (square), 6300\AA\ (asterisk), 6363\AA\ (diamond), uncorrected for extinction but corrected for the continuum variations. }%
    \end{figure}

The [O II] 4443.516\AA\ line was likely present , displaying  the combined extended wing and core (intermediate) profile.  [O II] 7319.92, 7330.19\AA\ were both present with nearly symmetric, broad profiles but with central wavelengths shifted by +23$\pm$2 km s$^{-1}$ from the system velocity.  The lines, fit with gaussian profiles, have FWHM$\approx$58$\pm$6 km s$^{-1}$.  Their profiles are the same as those underlying the narrower emission on the other oxygen lines.   These  lines were difficult to detect in the final 2010 spectra because of the increased contribution of the Mira continuum.  We suggest that the lower ionization potential for N$^+$ compared to O$^{+2}$ accounts for the profile differences between the species;   the ionization potential for O$^{+}$, 35 eV, exceeds that of N$^o$, 14.5 eV.
            
    \begin{figure}
   \centering
   \includegraphics[width=8cm]{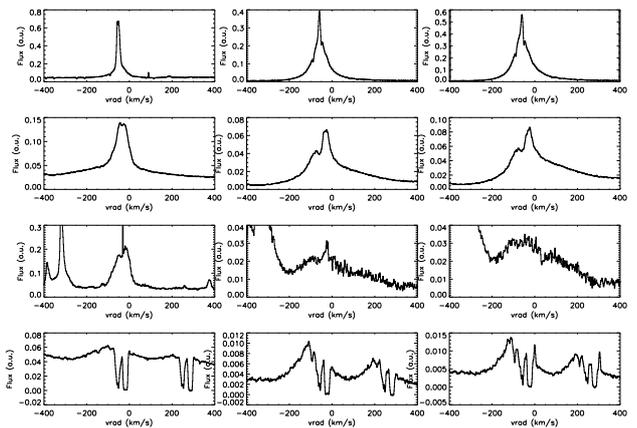}
   \caption{Comparative line profile variations of  [O I] 6300\AA\ (top), [N II] 5754\AA\ (second),  [O III] 4363\AA\ (third), and Na I 5889\AA\ (bottom) from the NOT sequence for JD  55288 (left column), JD 55334 (middle column), and JD 55350 (right column).  The separation of chromospheric, circumstellar, and shock emission is evident in the sequence (see text).}%
    \end{figure}

The O III 4959, 5007\AA\ lines were weakly present, as narrow emission features, in the first spectrum, JD 55270 (Day 2) but there was no detection of 4363\AA\ before JD 55287 (see Fig. 9).  While the nebular doublet displayed two-peaked emission (see Fig. 31, below), the 4363\AA\ profile did not and rapidly passed to the same form we found for the highest ionization lines (see Fig. 10).

 \begin{figure}
   \centering
   \includegraphics[width=8cm]{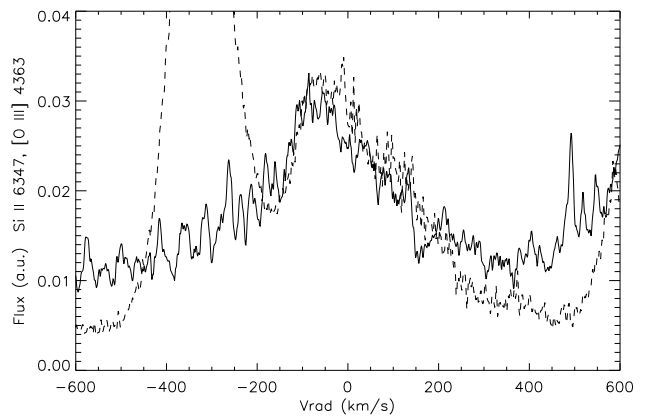}
   \caption{ Comparison  of the late-time [Fe X] 6374 \AA\ line profile (solid) with [O III] 4363 \AA\ (dash) on JD 55369 in the NOT sequence.  The spectra are scaled for superimposure.  The line may also be blended with Si II 6347 \AA\ and this comparison shows that the [O III] line was more symmetric than the usual shock features and extended to about 350 km s$^{-1}$ from the rest velocity.  }
    \end{figure}

%%%----   The most obvious difference between the O$^{+2}$ and  O$^o$ 

No N I lines were detected during the entire outburst.  The  [N II] 5754 \AA\ line was detected in the first NOT spectrum and showed some unusual behavior.  The profile was initially symmetric and comparatively narrow,  with a FWHM of 70 km s$^{-1}$ with shallow extended wings from -300 km s$^{-1}$ to +250 km s$^{-1}$ that matched those on the O I lines, but with the centroid velocity displaced to -35 km s$^{-1}$.   This last feature was confirmed on three spectra taken within the first two days of the NOT sequence (see Paper I for the journal of observations).  The first NOT spectrum, obtained with the highest S/N ratio,  shows [N II] to be double peaked, -49$\pm$1 and -29$\pm$1 km s$^{-1}$, with equal intensity components.  An intriguing feature of the comparison between [O I] 6300 \AA\ and [N II] is that the wings of the neutral oxygen line closely match the [N II] profile.  This indicates that the O I has separate contributions from  the chromosphere of the Mira, the narrow (FWHM $\approx$10 km s$^{-1}$) doublet at -54 km s$^{-1}$ and a second, environmental component, redward displaced relative to the chromospheric line.  The simplest explanation is obscuration of that part of the wind that is blueshifted relative to the observer.    The [O III] 4363 \AA\ line from the first NOT spectrum shows the same profile {\it and} velocity centroid as [N II] 5754 \AA, supporting this contention since it is not expected to have a chromospheric component.  The subsequent development of the [N II] is similar to the H$\delta$ line (see Fig. 19) for which individual emission peaks develop at the same velocities after JD 55350 and the wing is also similar.  As with all broad profiles, the extended wings are asymmetric (see Paper I).

    \begin{figure}
   \centering
   \includegraphics[width=9cm]{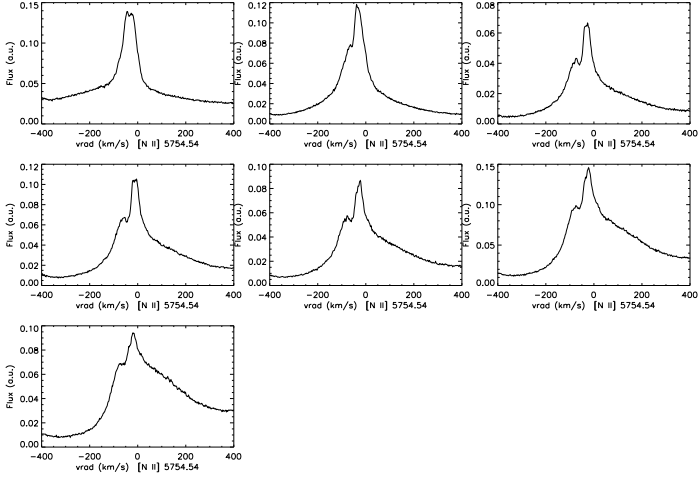}
   \caption{Evolution of the [N II] 5754\AA\ line profile from the NOT spectra for JD 55288, 55314, 55334, 55350, 55370, 55378, and 55397.}%
    \end{figure}

The emission fine structure on the [N II] nebular lines changed during the spectral sequence.  Three emission components were noted, especially on [N II] 6583 (see Fig. 11).  The relative strength of the component at -80 km s$^{-1}$ decreased but may also have been rendered less visible by the slightly lower resolution of the last two NOT spectra.  It occurred roughly at the velocity of the increasing absorption component on the Na I D lines (Paper I) .  Any feature on Na I D at -20 
km s$^{-1}$ was, instead, masked by the interstellar absorption component .  The ratio of the [N II] components at -42 km s$^{-1}$ and -20 km s$^{-1}$, however, changed dramatically and secularly.  Assuming that the decrease in the former was due to recombination, using the recombination coefficients from Osterbrock \& Ferland (2006) gives $n_e\approx 10^5$ cm$^{-3}$ for a timescale of about 80 days.  The relative constancy of the -20 km s$^{-1}$ line is consistent with a  lower 
%previously, higher
 density.  The inversion of the line ratio is also consistent with recombination of N$^{+2}$ to N$^+$ after the first NOT observation, at which time the [N II] lines were unobservable because of the strength and breadth of the H$\alpha$ line.
    
    \begin{figure}
   \centering
   \includegraphics[width=8cm]{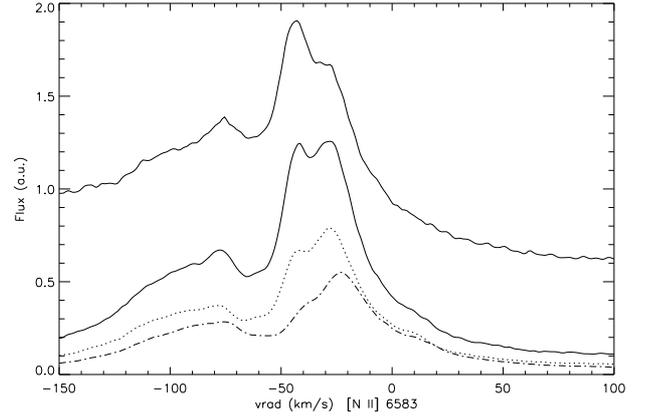}
   \caption{Variation in the low velocity emission line components of [N II] 6583\AA; from top: 55314 (solid), JD 55334 (dot), 55370 (dash), 55397 (dot-dash).  The emission hole at around -54 km s$^{-1}$, corresponding to the maximum in the narrow [O I] emission, remained constant.  This indicates the presence of  large density contrasts in the circumstellar material, comparing the -40 km s$^{-1}$ and -20 km s$^{-1}$ features; some weakening appeared also for the -80 km s$^{-1}$ feature.  These coincided with the emission/absorption components on Na I D (see text).}%
    \end{figure}
     
\begin{figure}
   \centering
   \includegraphics[width=9cm]{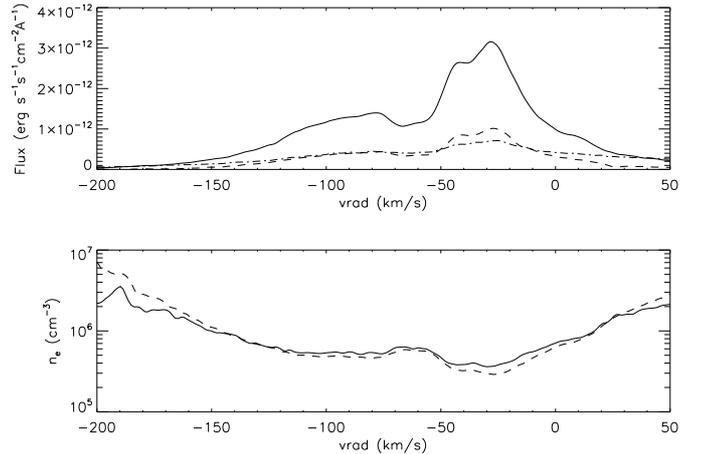}
   \caption{ Top: The three [N II] profiles for JD 55357, [N II] 5754 \AA\ (dot-dash), 6546 \AA\ (dash), 6583 \AA\ (solid).    Bottom: [N II] nebular diagnostic giving the electron density ($n_e$) as a function of radial velocity for T$_e$=10$^4$K for JD 55357 (solid) and JD 55369 (dash).   The flux was dereddened using E(B-V)=0.5 and the underlying continuum and H$\alpha$ emission wing were subtracted (see text).}   
    \end{figure}
    
As an independent determination of the electron density, we  were able to exploit the high resolution of the NOT spectra to study the spatial variation of the electron density using the [N II] nebular diagnostic ratio of [F(6583)+F(6548)]/F(5754) (Osterbrock \& Ferland 2006).  This is shown in Fig. 12 as a function of velocity.  The fluxes were dereddened using the Cardelli, Clayton, \& Mathis (1989) extinction curve with E(B-V)=0.5 (see Paper I) and the underlying continuum and wings of the H$\alpha$ line were subtracted.   Only the narrow components are shown, the broad emission from the shocked wind has an electron  density that was at least a factor of 10 higher.  The derived value for the electron density, a few times 10$^{5}$ cm$^{-3}$ agrees with that obtained from  the recombination timescale.  

Figure 14 compares the first NOT [O I] 6300\AA\ profile with [N II] 5754\AA\, showing that the narrow core chromospheric component was absent on the nitrogen line.  Figure 15 compares [N II] 6548\AA\ with He I 5875\AA\ highlighting the environmental component of on the helium line.
   
    \begin{figure}
   \centering
   \includegraphics[width=7cm]{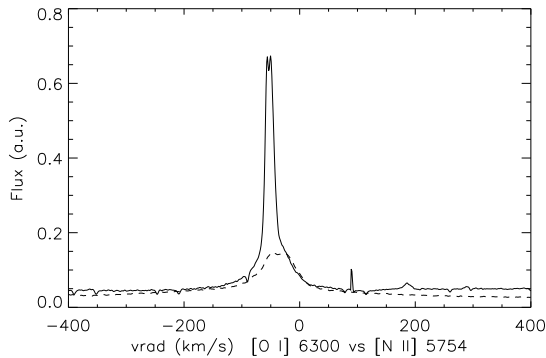}
   \caption{Comparison of [O I] 6300\AA\ (solid) and [N II] 5754\AA\ (dash) emission line profiles from JD 55287.}
    \end{figure}
   
    \begin{figure}
   \centering
   \includegraphics[width=9cm]{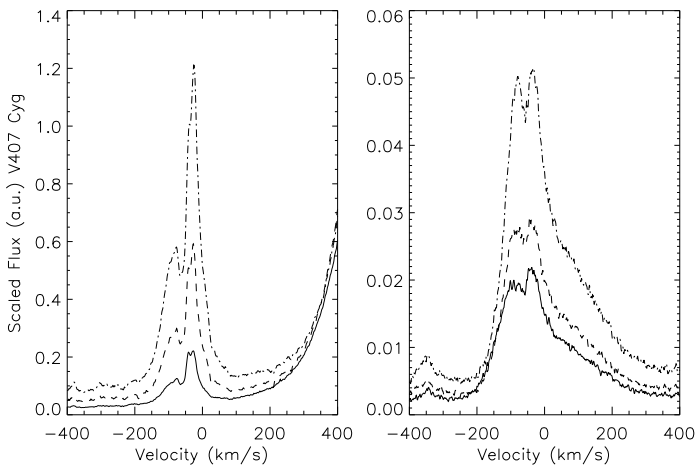}
   \caption{Comparative development of (left) [N II] 6548\AA\ and (right) He I 5875\AA\ line profiles for JD 55334 (solid), 55350 (dash), and 55370 (dot-dash).}
              \label{outburst-spectra}%
    \end{figure}
  
\subsubsection{Ne, Mg}

The [Ne III] 3960\AA\ line was present in the earliest NOT spectra, thereafter its detection was impossible because of low signal-to-noise ratio.  The profile was also severely distorted by the Ca II H absorption.   Thus, we can only say that Ne$^{+2}$ was present as early as two weeks after the outburst with a velocity width and profile that were similar to the [O III] lines on the same day.  The Ca II 3968\AA\ line lies, fortuitously, on the wing of the [Ne III]  3960\AA\ line and therefore the absorption components are more easily studied than on Ca II 3933\AA.  There was an extension of the emission that is consistent with a broad component (from -200 to +150 km s$^{-1}$), likely similar to Na I D, on the Ca II H line.  The absorption components approximately coincide in radial velocity with those on Na I  D.

Of the three lines in the resonance triplet of  Mg I 4562.60, 4571.10, 4575.29\AA, only  the 4571\AA\ line was detected; it has the largest transition probability, the companion lines are strongly forbidden M1 transitions).  As discussed in Paper I, its equivalent width is linearly correlated with [O I] 6300 \AA.  In Fig. 16 we display the profile variation during the NOT sequence.  The strong, narrow central component is likely chromospheric while the broad wing comes from the shock.  The broad profile is similar to that on the Na I D lines and the O I profiles with the same high velocity asymmetry.   The last panel shows the first NOT spectrum profiles for 5167\AA\ absorption line and 4175\AA\ emission line core in the velocity range from -100 to 0 km s$^{-1}$.  Note that the emission corresponded to the blueward absorption component of the 5167\AA\ profile and also agreed with the velocities of the 5172 and 5183\AA\ lines.

%$A$-value, 2.54$\times$10$^2$ s$^{-1}$,  
\begin{figure}
   \centering
   \includegraphics[width=8cm]{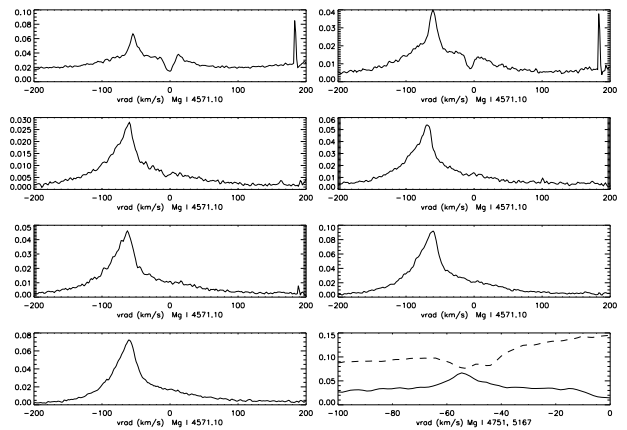}
   \caption{ Line profile development for Mg I 4571.10\AA\ line in the NOT sequence.  Note,  as with the development of [O I] 6300\AA\ intercombination line, the combined contributions of chromospheric and shock emission.  The dates are the same as those in Fig. 10 with the exception of the last panel.  This shows the emission peak at 4571\AA\ and the absorption at 5167\AA\ on JD 55287.  See text for discussion}
    \end{figure}

\subsubsection{Si, S}
    
The Si II 6347\AA\ line was overwhelmed by the coincident [Fe X] line that displayed a broad profile throughout the observing interval with wings extending to nearly 400 km s$^{-1}$ and no P Cyg structure at any time.  This Si II line was not detected on  JD 55674.  The Si II 5041\AA\ profile was broad in the first NOT spectrum,  HWZI $\approx$400 km s$^{-1}$, but displayed an unusual subsequent evolution due to a blend at 5043.3\AA.  The Si II 5056\AA, instead, showed that -- as with all excited state transitions -- the profile was similar to He II 4686\AA, displaying the resolved structure in the NOT spectra  at the same velocities (see Paper I).  The nearly perfect wavelength coincidence between Si II 5041\AA\ and [Ni IV] 5041\AA\ may account for some of the Si II profile evolution.  Without the confirming 5056\AA\ profile variations these could not have been disentangled.  The profiles were almost identical to [O III] 4363\AA\ with the structure on the different Si II profiles being caused by blending with different superimposed Fe II emission lines.   
    
    \begin{figure}
   \centering
   \includegraphics[width=9cm]{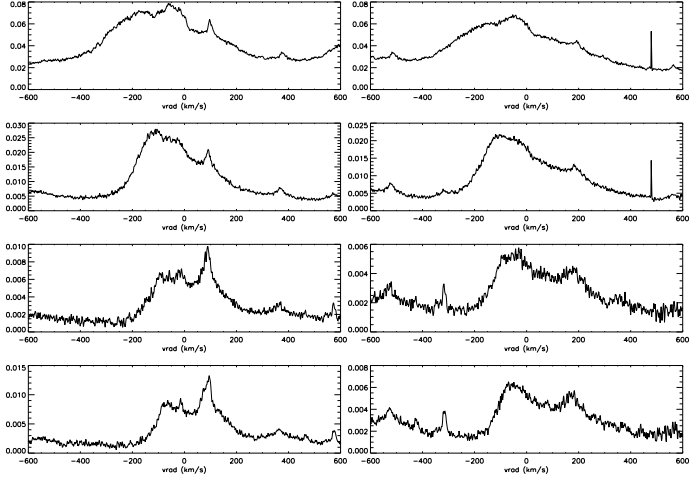}
   \caption{ Left: Si II 5041 \AA\ from JD 55288, 55314, 55370, and 55397.  The line is blended with [Ni IV] 5041\AA. Right: Si II 5056\AA\ on the same days.  This comparison also demonstrates that the contamination of the 5041\AA\ line is small (see text).}
    \end{figure}
   
The [S II] 4068, 4075 \AA\ lines remained comparatively narrow throughout the observing sequence but never developed the double peaked structure observed in the low density tracers.  The development of the [S II] 6716, 6730\AA\ doublet  was described in Paper I but the blue doublet was not discussed there.   The FWHM increased from 55 km s$^{-1}$ on JD 55334 to 92 km s$^{-1}$ on JD 55397.  In the first two observations, the [S II] 4068 \AA\ line was blended on the wing of a much stronger Fe II emission line.    The [S III] 6312.10 \AA\ line was present from the start of our observing sequence, displaying the same profile as [O III] 4959, 5007\AA\ and [Ar III] 7135\AA\ after JD 55288 in the higher resolution NOT spectra, unlike the [S II] 6730\AA\ line discussed in Paper I that traces the shock.  This line was still detected on  JD 55674.  The line is a tracer of the thermal emission from the post-shocked gas in the expanding ejecta and the recombining Mira wind.  We show the equivalent width variations in Fig. 18, the line measurements have been corrected for the continuum variations listed in Table 2.  We found that this line also varies in the same way as the radio continuum measurements.  Centimeter wavelength observations began with OCRA at the announcement of the outburst and continued throughout the observing period (Peel et al. 2010).\footnote{The data are available at:\\ http://www.jb.man.ac.uk/research/ocra/v407cyg/ }  The star was first detected in the radio region with an inverted spectrum, S$_\nu \sim \nu^{0.97}$, on  JD 55278 (Nestoras et al. 2010), indicating that the ejecta (and surrounding environment)  were still optically thick.  The spectrum gradually flattened (S$_\nu \approx {\rm constant}$ after JD 55350),\footnote{See the EVLA site, https://safe.nrao.edu/evla/nova/\#v407cyg,\\ and Krauss et al. 2010.} likely displaying the transition to an optically thin gas.  The peak of radio emission at all wavelengths coincided with the strongest [S III] 6312\AA\ emission.  Thus the radio in the earlier stages could have been from the shock while later, it was from bremsstrahlung cooling of the ionized Mira wind.

  \begin{figure}
   \centering
   \includegraphics[width=8cm]{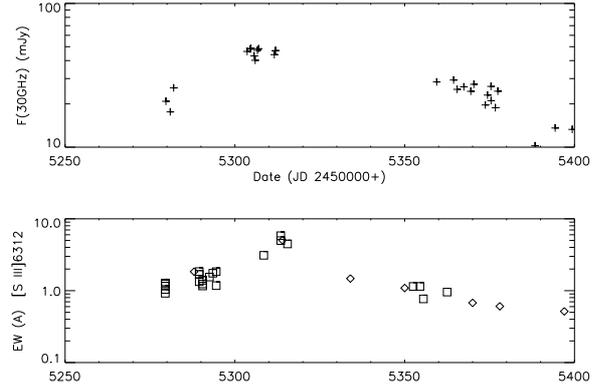}
   \caption{Top: 30 GHz flux from OCRA (see text); Bottom: [S III] 6312\AA\ equivalent widths corrected for the continuum contribution: Square: Ond\v{r}ejov; triangle: NOT.  See text for further discussion.}
    \end{figure}
    \subsubsection{Ar, Ca}

The [Ar III] 7135.79\AA\ line resembled the [N II] 6548, 6583\AA\ lines and the ground state [O III]  4959, 5007\AA\ lines, with two different contributors to the profile.  One was the broad component we associated with the shocked wind and the expanding ejecta.  The other was the low density region that had an expansion velocity of about 70 km s$^{-1}$.  This region could have been produced by the symbiotic nova eruption in the 1930s at a distance of about 2900 AU (about 0.02 pc); this is an upper limit based on the appearance of the profile over the 26 day interval between the first and second NOT spectrum.  If we assume that the ejection lasted a few years, as indicated by the light curve, the ejecta thickness would be about $\Delta R/R \approx 0.15$ and with a radial expansion law of  $R(t)\sim t^{2/3}$  would reach this distance in the $\approx$70 years after the end of the event.    With a density of $3\times 10^4$cm$^{-3}$ , the mass of the ionized region is about 10$^{-2}$M$_\odot$ implying the mass loss rate during the symbiotic nova event to be about 10$^{-3}$M$_\odot$yr$^{-1}$.  This is, again, an upper limit assuming that the event lasted as long as the visible maximum.  The line we had previously ascribed in Paper I to [Ar IV] 4711\AA\ line is He I 4713\AA.  

The resonance line of Ca I, 4226.73\AA\ was present in absorption at a velocity of -55.2$\pm$1.5 km s$^{-1}$ through JD 55314 (but the last had a low signal-to-noise ratio).  Thereafter,  mainly because of the weak exposure at 4200\AA\, it was not detected.  Neither  [Ca I] 4575, 4912, 4916\AA, nor the weak 6572.78 \AA\ resonance line were detected at any time.   The Ca II 3969 \AA\ line showed the same components as Na I D, the 3933\AA\ line had a weaker absorption at the  rest velocity on JD 55288 but was severely blended with H$\epsilon$.  The [Ca II] 7291, 7323\AA\ doublet lines were strong and extremely narrow, see Fig. 20,  and coincident in velocity with [O I] 6300\AA.   The Ca II infrared $^2$D-$^2$P$^o$ multiplet, 8498.02, 8542.09, and 8662.14\AA, is shown in Fig. 19.  As can be seen in the figure, the profiles were the same for all three lines.  The wings were quite extended, similar to the permitted Fe-peak lines and with similar profiles, throughout the 2010 observing period and a weak absorption at the save velocity as the metallic lines and Na I D appears to have been present in the earliest spectra.  These lines are pumped by Ca II H and K.  
    \begin{figure}
   \centering
   \includegraphics[width=8cm]{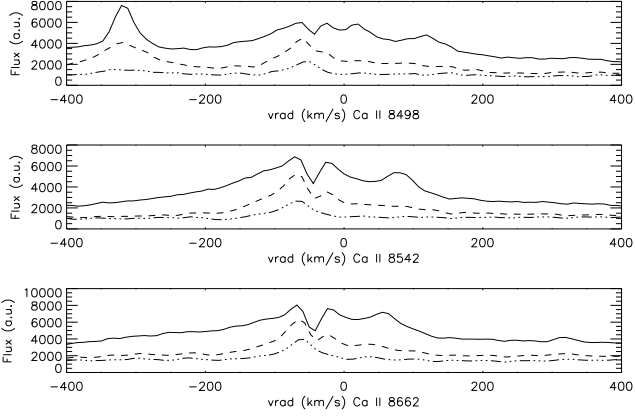}
   \caption{Evolution of the Ca II triplet components in the Ond\v{r}ejov spectra from JD 55279 (solid),   JD 55315.50  (dash);  JD 55362 (dot-dot-dash).   The flux units are arbitrary, the profiles are scaled for display.}%
    \end{figure}

    \begin{figure}
   \centering
   \includegraphics[width=9cm]{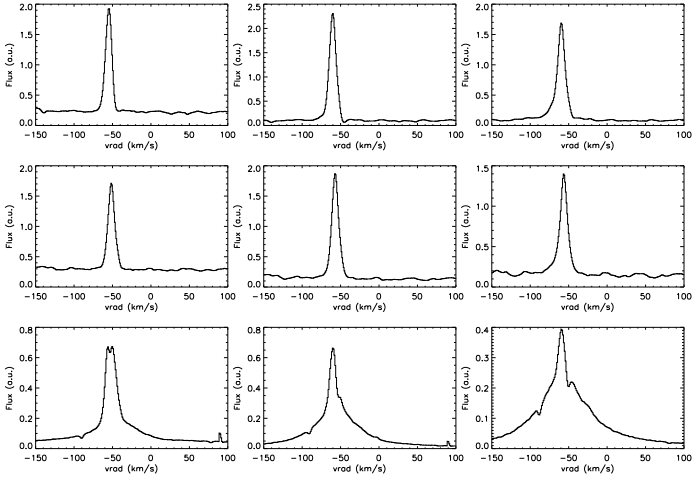}
   \caption{Comparison of chromospheric Ca II 7291\AA\ (top), 7323\AA\ (middle), and [O I] 6300\AA\ (bottom) line profiles for the first three NOT spectra, JD 55288, JD 55314, and JD 55334.}
    \end{figure}
         
\subsection{Neutral and singly-ionized Fe-peak emission lines}

The most complex profile development was observed among the iron peak transitions.  There appear to have been distinct mechanisms: fluorescence, chromospheric thermal emission, ionized wind recombination, and collisionally excited post-shocked gas.  The first, due to the strong coupling with ultraviolet transitions, dominated the narrow line emission spectrum through JD 55314.  In this interval, many of the strongestpermitted  transitions of Cr II, Ti II, and Fe II showed P Cyg or isolated absorption profiles.   These persisted on only the most strongly pumped multiplets such as Fe II RMT 42 (see Fig. 21).    The high optical depth of the wind and chromosphere, exposed to the UV from the WD and the shock, produced a rich fluorescence spectrum among the Fe-peak species.   These will be discussed in detail in the next paper in this series (Wahlgren et al., in preparation).  In general, the line profiles could be decomposed into three contributors that varied independently.  The chromospheric emission, as observed for the [O I] and [Ca II], was a narrow component with FWHM of about 10 to 15 km s$^{-1}$ centered at the rest velocity of the Mira.  This was at the same velocity as the isolated absorption features.  For instance, the Fe II 5018\AA\  P Cyg absorption component was at -60$\pm$1 km s$^{-1}$, the same velocity as minimum of the H$\gamma$ absorption and on the blueward side of the [O I] 6300\AA\ emission peak.

 \begin{figure}
   \centering
   \includegraphics[width=9cm]{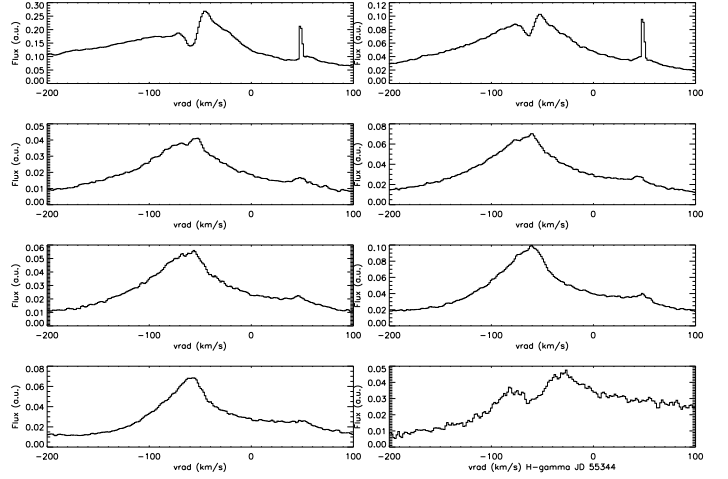}
   \caption{Example of the Fe-peak P Cyg absorption and emission variations.  Fe II 5018\AA\ for the NOT 2010 sequence.  The order of the spectra in time is the same as in Fig. 11; the last panel -- lower right --  shows the H$\gamma$ line from JD 55334 for comparison with the low velocity P Cyg absorption trough. }
    \end{figure}

There was possibly emission from [Cr III] 5823\AA\ (but not 5600\AA, 541 \AA) on JD 55369; the feature strengthens in the later spectra but provides little profile information. 

The variations of the strongest [Ni II] transitions are shown in Fig. 22.   The [Ni II] 6666.8, 4326.24\AA\ lines displayed similar, narrow profiles throughout the NOT 2010 sequence centered on the system velocity.  The small shift in the 4326\AA\ line was due to a blend with [Ni III] 4326.16\AA.  The excited state line 6813.57\AA\ was present but in the later NOT sequence it  was superimposed on the strong Mira continuum so only the first spectrum reveals the nebular component.   The [Ni III] 6002.2\AA\ line was unblended and also narrow but consistently weak.  The strong  [Ni IV] 5041.56\AA\ was superimposed on Si II 5041\AA\ and was unmeasurable.
    \begin{figure}
   \centering
   \includegraphics[width=9cm]{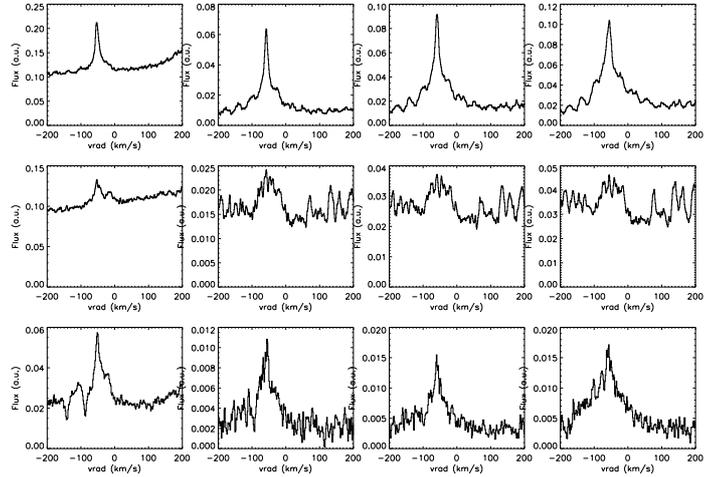}
   \caption{Comparison of [Ni II] 6666.80\AA\ (top) and  6813.57\AA\ (middle), and 4326.24\AA\ (bottom) for JD 55288, JD 55334, JD 55370, and JD 55397.}
    \end{figure}
    
\subsection{Narrow emission lines from the ionized wind}
          
The first NOT spectrum (JD  55286) showed identical profiles on all of the [O III], [Ar III], and [N II] lines.  After JD 55370 the [O III] 4363\AA\ and [N II] 5754\AA\  lines developed shock (i.e. He II and [Ca V]) profiles.   The [O III]  4959, 5007\AA\  lines in the first NOT spectrum displayed multiple peaks, confirmed by independent spectra on the same day, at -130, -67, -39, and -21 km s$^{-1}$ (with an uncertainty of $\pm$5 km s$^{-1}$).  These emission peaks are actually  paired relative to the Mira velocity that we obtained from the Li I 6707\AA\ photospheric line and appear to arise from separate shells with expansion velocities of 76 km s$^{-1}$ and 13 km s$^{-1}$.  The higher velocity is greater than that observed  for the absorption in the Na I D line that we reported increasing in the  later observations  after the XR peak.  The minimum in the line profile at -55 km s$^{-1}$ coincides with the emission peak of  [O I] 6300, 6363\AA\ and the strongest stellar absorption feature on the Na I D lines.   As noted in Paper I, the Na I absorption at -95 km s$^{-1}$ steadily increased following the XR peak and this absorption coincided  with a local minimum in intensity in the [O III] emission line profiles.  The corresponding emission feature noted in the spectra after day 60, at 0 km s$^{-1}$, had no counterpart on either the [O III] or [O I] profiles but in the later spectra coincided with the broad redward wing.  The Na I D high velocity absorption had, however, a correspondence with a weak emission peak on [O I] 6300\AA\ at -92 km s$^{-1}$.  This comparison suggests that the features  displayed on these low ionization line profiles probed a structure at two levels of the Mira environment.  The lowest velocity components, relative to the system velocity, are likely chromospheric.  Those at higher velocity could then be due to shells surrounding the system expelled at velocities of order 75$\pm$5 km s$^{-1}$.  The low ratio of the blueshifted to redshifted components of the [N II], [O III], and [Ar III] lines, compared with the nearly symmetric neutral lines, suggests that the Mira was interposed along our line of sight and  shadowing the white dwarf and the inner portion of the expanding shocked wind.  The [O II] 7319\AA\ and 7330\AA\ equivalent width variations are shown in Fig. 23.

The comparison of the isoelectronic transitions of [N II] and [O III] shows the different contributions of the circumstellar and shock emission.  Completing  the C I   isoelectronic sequence, the only low excitation carbon line in the visible spectrum, [C I] 4621.57\AA, was not detected.  The narrow [O I] component corresponds, in velocity, to the gap in the ion line emission profiles.  All of the profiles have, however, high velocity wings that extend to at least 200 km s$^{-1}$ from the Mira velocity and in the excited state transitions of [O III] and [N II] are asymmetric in the same sense as the He II and [Ca V]  shock lines we discussed above.  

\begin{figure}
   \centering
   \includegraphics[width=8cm]{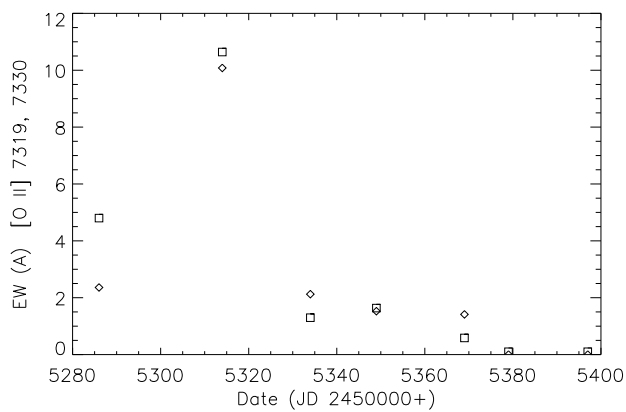}
   \caption{ Equivalent widths (\AA) of [O II] 7319\AA\ (square) and 7330\AA\ (diamond) from the NOT sequence, corrected for continuum contribution from the shock (see text).}
    \end{figure}

 \begin{figure}
   \centering
   \includegraphics[width=8cm]{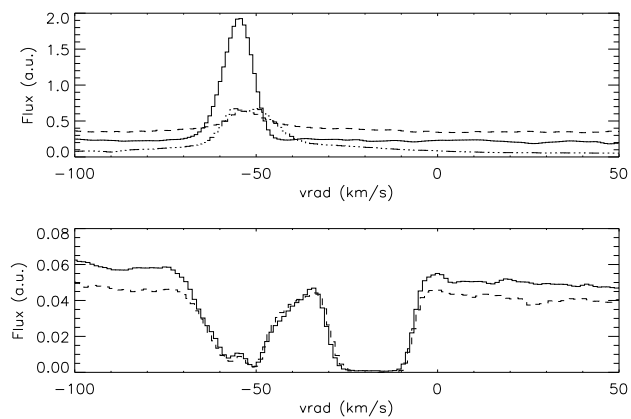}
   \caption{Comparison of narrow line profiles from the first NOT observation, JD 55288, showing the narrow component before the development of the extended wings (day 23 after outburst).  Note that the central peak in the Na I D absorption line coincided with the peak of the other emission lines, suggesting that this was also a superimposed chromospheric emission component.   Top: [Ca II] 7291\AA\ (solid), [O I] 6300\AA\ (dash), [Ni II] 6666\AA\ (dot-dot-dash).  Bottom: Na I 5889\AA\ (solid), 5895\AA\ (dash) for comparison.}
    \end{figure}

    \section{The aftermath: 2011}
    
     \begin{figure}
   \centering
   \includegraphics[width=8cm]{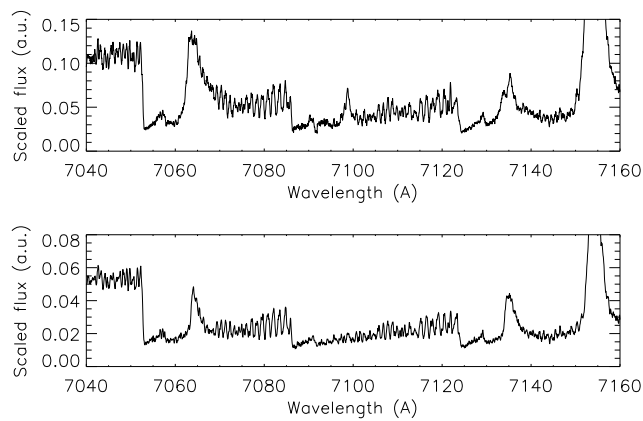}
   \caption{Comparison of  Mira continuum and far red emission lines from JD 55397 (top) and JD 55794 (bottom).  The strong line at the red end is [Ar III] 7135\AA. }
    \end{figure}
    \subsection{The red giant spectrum}
    
    Thus far we have discussed the phenomenology during the first roughly half year of the outburst, through the end of the X-ray observations.  We now turn to the more recent developments as the system returns to its quiescent state.  We note, however, that based on the spectra shown by Munari \& Zwitter (2002), V407 Cyg is {\it not}  near its base state.  The presence of high velocity wings on most of the emission lines  shows that there is still expanding shocked gas dominating the spectrum formation shortward of 7000\AA, albeit at lower excitation than many of the species present during the preceding year.   The observations in 2011 were obtained during the next half-cycle of the Mira, on approach to maximum light, and thus with a larger radius for the giant.    The Li I 6707\AA\ line was measurable with an equivalent width of 400$\pm$20 m\AA, essentially unchanged from the late 2010 spectra and indicating that the veiling continuum was completely gone.  A sample region is shown in Fig. 25, a region that our calibration spectrum of BD +28$^o$4211 shows to be free of terrestrial contamination.  The He I 7065\AA\ line was still detectable but  nearly overwhelmed by the absorption bands of the Mira.   The most evident change is the disappearance of the weak emission feature at 7098.6\AA\ that was first detected on JD 55334 and persisted through the 2010 observing period.   This may be due to the increasing visibility of the Mira as the veiling continuum faded.  Its subsequent disappearance may be an opacity effect, that the increase in the band strengths and changing excitation at a different pulsational phase of the Mira removed this {\it pseudo}-emission gap.  The most sensitive parts of the continuum, those showing the VO and ZrO bands, were not, unfortunately, observed in 2011 and the TiO bands are less sensitive to changes in the spectral type of the giant.

     \begin{figure}
   \centering
   \includegraphics[width=8cm]{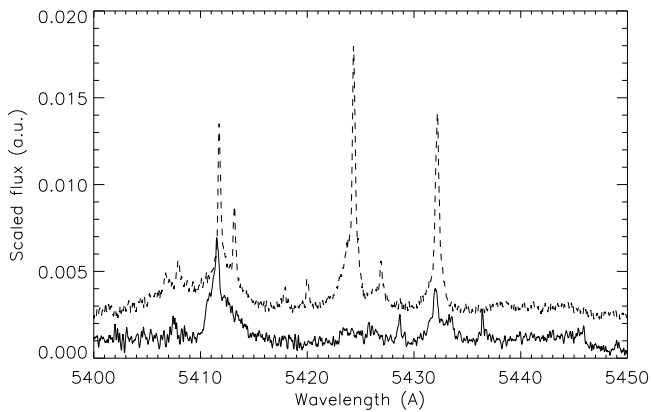}
   \caption{Comparison of  emission lines in a representative portion of the spectrum from the 2010 Apr. 2 (dash) and 2011 Aug. 21 (solid) observations.  Every persistent line was a forbidden transition, e.g. [Fe II] 5412, Fe II 5433\AA; the permitted lines were all systematically significantly weakened or absent in the later spectrum, e.g. Cr II 5407, Fe II 5414, Fe II 5425, Ti II 5421, Fe II 5427\AA.   See text for discussion.}
    \end{figure}
    
    \subsection{The emission lines in the late spectrum}
    
The emission line spectrum observed in the JD 55794 NOT data is so different from the previous developments that we here provide a separate description.   During the interim between the NOT observation, 2010 Jul 15, and this last one, all permitted transition lines of the Fe-peak that had dominated the emission spectrum in 2010 disappeared.  In every case, the lines that remained visible were forbidden (both M1 and E2) and mainly from [Fe II]).   These were composite in form, a narrow central peak near the rest velocity of the Mira and a broad, asymmetric profile with a maximum redward extension of 250 km s$^{-1}$, see Fig. 26.   One exception should be noted.  The [Fe II] 5412.65\AA\ line was strong and persistent but displayed an extended wing toward -200 km s$^{-1}$.  This line was a blend with He II 5411.52\AA, thus accounting for the unusual profile.  Although the signal-to-noise ratio is comparatively low in the last spectrum, for the strongest lines the central velocity of the narrow component was -60 km s$^{-1}$.  The strong lines in the JD 55674 Ond\v{r}ejov spectrum all show identical profiles to those of the convolved NOT spectrum.  There was only a weak blueward asymmetry of the H$\alpha$ peak and the same [N II] profiles.  The weaker emission lines, detected in the NOT spectrum, were not visible but the absence of the stronger permitted transitions of the Fe-peak lines, and the persistence of the forbidden transitions, was confirmed.  
    
    \begin{figure}
   \centering
   \includegraphics[width=8cm]{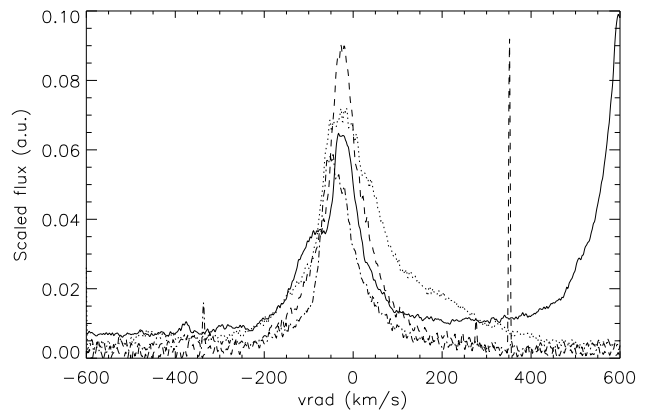}
   \caption{Comparison of the [N II]  6548\AA\ (solid), [N II] 5754\AA\ (dot), H$\beta$ (dash), He I 5875\AA\ (dot-dash) lines from JD 55397.}
    \end{figure}
    
 \begin{figure}
   \centering
   \includegraphics[width=6.5cm,angle=90]{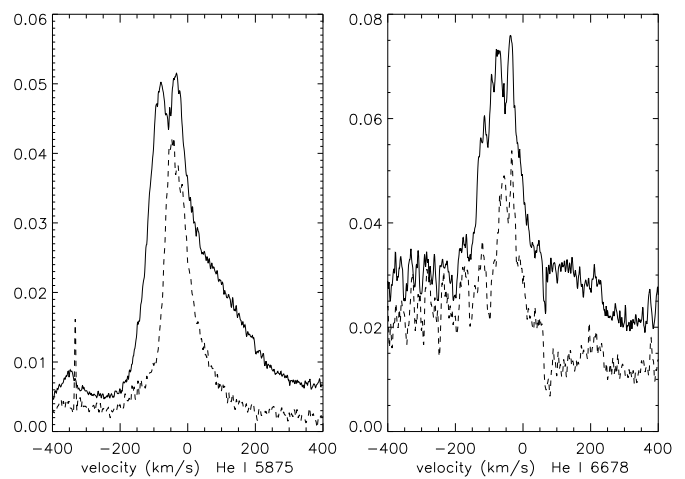}
   \caption{Comparison of the He I 5875\AA\ (left) and 6678\AA\ (right) lines from JD 55397 (Day 132, solid) and 55794 (Day 529, dash).  Much of the structure on the 6678\AA\ line is from the underlying continuum absorption of the Mira.}
    \end{figure}
    
We display in Figs. 27 through 32 galleries of the nebular emission line development from the NOT observations.  In all transitions, there were two separate changes.  The red wing increased, most notably for [N II] 5754\AA\ and  [O III] 4959, 5007\AA, and the blueward peak at -100 km s$^{-1}$ significantly diminished in contrast.   The [Ar III] 7135\AA\ line showed an even more evident decrease in the blueward peak but the Mira continuum was stronger and the profile details are less secure.    The  blueward emission peak on He I 5875\AA\ also vanished but, in contrast, the redward extended emission {\it decreased}.   The wings on [O I] 6300\AA\ that were already present by JD 55397 remained constant and the final profile resembled the persistent forbidden Fe-peak lines with a similar velocity and FWHM for the narrow component and a similar red wing.   The broad emission profile that had previously been present on the Na I D lines was not detected although the absorption component from the wind remained.  The signal-to-noise and contamination with terrestrial emission prevents a more quantitative statement.   We postpone the use of these different profile templates for decomposition of the line components to the next section.
     
 \begin{figure}
   \centering
   \includegraphics[width=8cm]{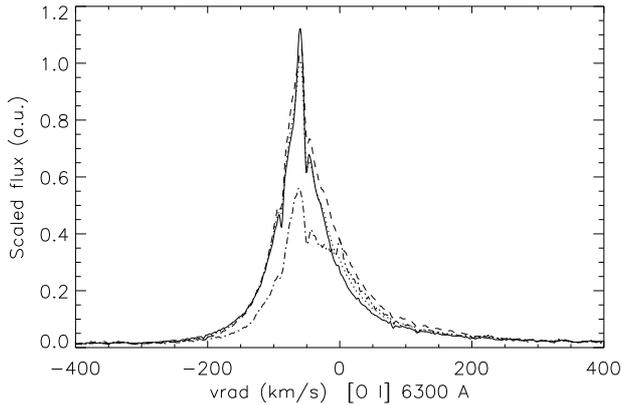}
   \caption{Comparison of the [O I] 6300\AA\ line from the NOT spectra from 2010 and 2011.  Solid: JD 55350; dot: 55370; dash: 55397; dot-dash: 55794.  The spectra have been normalized to the continuum of the Mira in all cases.  The weak fine structure is due to terrestrial water vapor contamination.}
    \end{figure}

 \begin{figure}
   \centering
   \includegraphics[width=8cm]{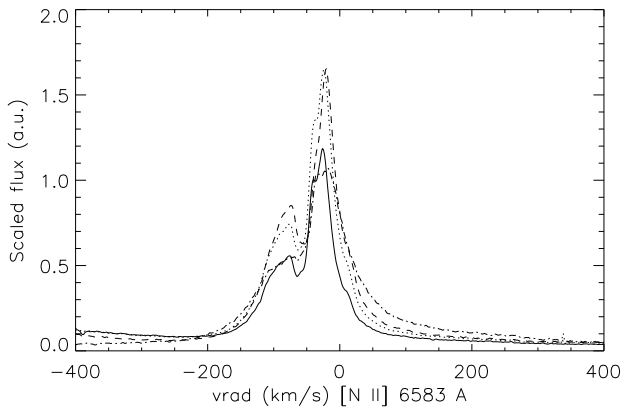}
   \caption{As in Fig. 29, the development of the [N II] 6583\AA\ line.  The [N II] 6548\AA\, as shown in Fig. 27, is blended on the redward side with H$\alpha$ but the same profile variations are evident.}
    \end{figure}

 \begin{figure}
   \centering
   \includegraphics[width=8cm]{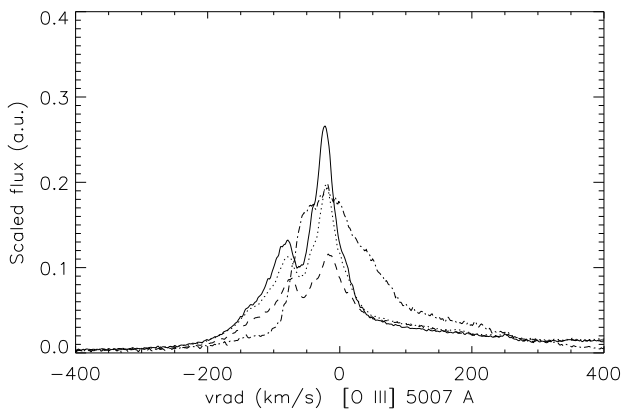}
   \caption{As in Fig. 29 for the [O III] 5007\AA\ line profile}
    \end{figure}

 \begin{figure}
   \centering
   \includegraphics[width=8cm]{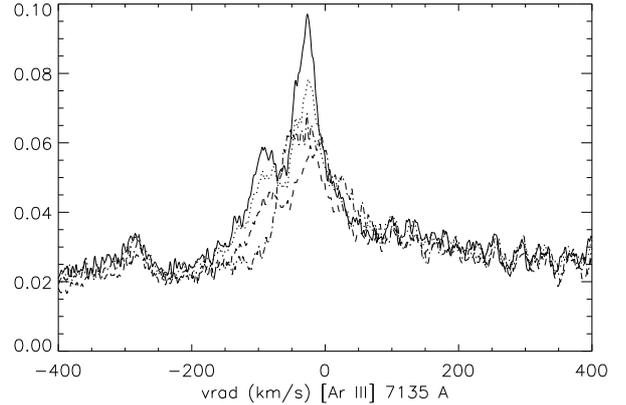}
   \caption{As in Fig.29 for the [Ar III] 7135\AA\ line profile}
    \end{figure}
    
In summary, sometime between the end of the 2010 observations and the 2011 Apr. 23 spectrum, it appears that the receding portion of the shock had expanded to nearly reach the line forming region for the narrow nebular transitions.  The approaching portion, on the other hand, appears to have stalled and the near side had probably recombined.

\section{Discussion: the circumstellar environment  of the V407 Cyg system}

Among the known symbiotic stars, V407 Cyg is unique.  Even compared with the limited sample of the symbiotic-like recurrent nova systems, the V407 Cyg system is unusual (Anupama 2008, Mikolajewska 2008, Bode 2010).  Among classical symbiotic systems, its Mira is the longest period pulsator known (Munari et al. 1990, Belczy\'nski et al. 2002).   The progenitor of the Mira must have been massive, $>$4 M$_\odot$, based on the lithium excess (Tatarnikova  et al., 2003a,b; Paper I;  Munari et al. 2011).  Only T CrB and RS Oph, both SyRNe, show significantly enhanced $^7$Li (Wallerstein et al. 2008), reinforcing the identification of V407 Cyg as a member of this class notwithstanding that the 2010 outburst is the first {\it known} such event in this system.  It is the only SyRN system to show maser emission (Deguchi et al. 2011), the individual components of which varied during the outburst.  But perhaps most unusual was the event that led to its being labeled ``nova-like'' in the literature, a symbiotic nova-like event in 1936 that lasted for about five years during which time the visual brightness of the system increased by more than a factor of 10, a behavior that is {\it not} seen historically among the other SyRNe (in addition to T CrB and RS Oph, neither V745 Sco nor V3890 Sgr appear to have such variability).  There were no contemporaneous spectroscopic observations during the event of the 1930s but, if it were like other systems, there would have been continuous mass loss through a wind from the white dwarf induced by the onset of a thermonuclear envelope process following a low level of accretion.\footnote{The historical record is quite poorly sampled before 1990 and we know very little of what V407 Cyg was doing between the outburst of the 1930s and the beginning of photometric and spectroscopic monitoring in the 1990s.  The unanticipated explosion of V407 Cyg means it was not included in the photographic survey by Schaefer et al. (2009) of Galactic recurrent novae.}

The nova explosion produced a rich variety of phenomena, as we have detailed, that distinguish this outburst from other SyRNe.  The first indication of [O III] 4959, 5007\AA\ was on JD 55272 (Day 6) ; [N II]  6548, 6583\AA\  only appeared after JD  55289 (Day 24).  The narrow [O I] 6300, 6363\AA\ and Mg I 4671\AA\ emission lines were, in contrast, clearly visible from Day 4.   In the Day 4 spectrum, narrow emission lines from permitted Fe-peak ions were seen without the underlying broad wings; those appeared by Day 6 and had become asymmetric in opposite senses (as we discussed above) by Day 13.   The first appearance of the double peak emission on [O III] 4959\AA\ can be estimated from the comparison of the lower resolution Ond\v{r}ejov spectrum on JD 55292 and the first NOT spectrum on JD 55288.  This gives a limit to the radius of the ionized region of $\leq$2$\times$10$^{16}$ cm.  Assuming that this is matter ejected during the outburst in the 1930s, the width is $\Delta R/R \leq 0.15$ based on the duration of the event (about 5 yr)  and using the electron density from the [N II] analysis, $n_e \approx 3\times$ 10$^4$ cm$^{-3}$,  the ionized mass sampled by the nova line profile is $\leq$10$^{-4}$ M$_\odot$.  If this is due to a wind during a symbiotic nova-like event, the inferred mass loss rate was  $\leq 2\times$ 10$^{-5}$ M$_\odot$yr$^{-1}$ independent of the distance.   The density is lower than, but roughly compatible with, the electron density inferred from the recombination analysis of the Na I D lines at this velocity discussed in Paper I, about 10$^5$ cm$^{-3}$.  The disappearance of the blue component can be interpreted as recombination in a zone located on the side of the giant toward the observer and shadowed from the WD.  The redward extension would be, in this picture, the low density periphery that is still illuminated by the WD and also collisionally excited by the shock.  The similar maximum positive radial velocity on all the forbidden transitions and the absence of the permitted lines, point to the emission arising in a very low density region.  The distinct taxonomic separation of the line profile asymmetry between permitted and forbidden transitions supports the picture that the blue wing of the E2 and M1 lines is suppressed by higher densities being encountered in the approaching part of the shock. and the disappearance of this component on almost all of the strong lines in the last spectrum being due to recombination.    The gap between the 2010 and 2011 observing sessions is, however, too long to constrain the densities.

 \begin{figure}
   \centering
   \includegraphics[width=8cm]{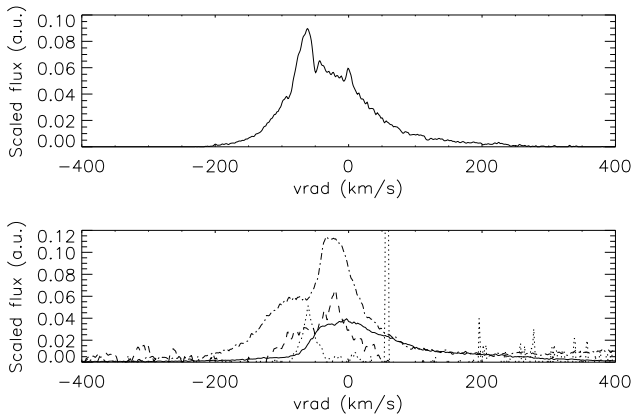}
   \caption{An example of the separation of the various contributors to the observed line profiles in the 2011 Aug. 21 NOT spectrum.  Top: [O I] 6300\AA.  Bottom: [O III] 5007\AA\ (solid), Mg I 4571\AA\ (dot), [S II] 6716\AA\ (dash), [N II] 6548\AA\ (dot-dash).  The Mira continuum has been removed in each but the scaling is for display (arbitrary flux units).  The spike at +60 km s$^{-1}$ is a cosmic ray hit near Mg I.   See text for description.}
    \end{figure}

The late-epoch spectrum from JD 55794 further clarifies the various contributions to the profiles.  As an example, we used the [O I] 6300\AA\ line (which was identical in profile to the related 6363\AA\ line) for the decomposition, shown in the upper panel of Fig. 33.  In the lower panel of Fig. 33 we show an empirical decomposition using a set of representative profiles from the data itself with the continuum of the Mira subtracted.  The relative intensities are chosen for display, they are not results from a fitting procedure.   The Mg II line was narrow with a FWHM of 25 km s$^{-1}$ centered at -61$\pm$1 km s$^{-1}$, the same radial velocity previously derived for the Mira wind absorption lines.   The [N II] 6548\AA\ showed an extended red wing to +200 km s$^{-1}$ and the persistent red peak.  This latter emission peak coincided with one on the [S II] 6716 and 6730\AA\ lines that also showed the reduced blue peak, like [Ar III] 7135 (Fig. 32).   The [O III] 4959\AA\ and 5007\AA\ lines no longer displayed the low velocity emission peaks, instead having a pure shock profile similar to [Ca V] 5309\AA.  The comparison indicates,however,  that a part of the [O I] profile is still being contributed by the part of the wind that has not been completely ionized and, as we proposed in Paper 1, the wing may be due to charge exchange behind the shock.ÊThe [S II] and [N II] lines still showed  contributions from the extended environment.  

Deguchi et al. (2011) present the variations of the SiO maser during the first month of outburst.  Taking $V_{Mira,LSR} = -26$ km s$^{-1}$, the main SiO  (J=1-0, v=2) component is at -30 km s$^{-1}$ (LSR), corresponding in heliocentric $v_{rad}$ to -58 km s$^{-1}$,  at slightly higher velocity ($\approx$ 4 km s$^{-1}$) than the peak of the [O I] 6300\AA\ line on JD 55286 and the mean absorption line velocity but compatible with the wind line velocities.
\footnote{There is an ambiguity here.  Deguchi et al. give the  system heliocentric radial velocity of -41 km s$^{-1}$, and discuss orbital motion with a 43 year period  citing the radial velocity from Tatarnikova et al. (2003b) using the Ca I 6573\AA\ line.  We detected this line in the last NOT spectra from 2010 and in the 2011 spectrum at $v_{\rm rad}$=-43 km s$^{-1}$.  There was, as we mentioned above, a weak absorption feature on Na I D at a similar velocity and a weak absorption component on Mg I 5167\AA.  The velocity disagreed, however, with all other absorption line values and we have no explanation for this other than either a blend or emission filling in the blue wing.  We adopt the velocity determined from the Li I 6707\AA\ and [Ca II] 7291, 7323\AA\ lines in our comparison.}    The (J=1-0, v=1) transition also shows a weaker line at -47 km s$^{-1}$ that does not corresponding to any absorption on the Na I D lines.   The other SiO maser line (J=1-0, v=2) shows only the -58 km s$^{-1}$ line.  The velocity is consistent with that of the Mira wind as determined from the absorption features discussed in the previous sections.  The rapid turn-off of the maser after the explosion, if due to the irradiation by the initial shock and the advance of the ionized region as a precursor, indicates that the cold material is located at a distance of about 1 to 10 AU from the Mira, consistent with the distance in Deguchi et al (2011).  
%We should note that the timescale for the disappearance of the wind absorption lines, $\leq$45 days, is longer than the timescale of about 20 days on which the maser lines turned on again. 

The orientation proposed by Deguchi et al. (2011) from the maser emission places the WD near quadrature at the time of the nova explosion.  The chromospheric lines, and the ``emission hole'' in the forbidden lines, are more consistent with a phase closer to superior conjunction for the WD -- that is, being seen through a substantial portion of the Mira wind.  Two results point to this.  The absorption line spectrum during the early outburst was strong, both on P Cyg components and in isolated absorption lines, and are more consistent with a long pathlength.  The other is the asymmetry of the high velocity component of the lines.  Were the WD seen nearer quadrature, it would not be possible to produce so asymmetric a shock line profile.   The persistence of the neutral lines also support the picture that the inner portion of the wind was screened from the UV of the explosion and shock while the bulk of the wind was ionized within the first 60 days.  If the fluorescent emission was excited by the UV of the WD, the disappearance of those transitions could indicate either that the ionization had removed the ions (mainly singly ionized Fe-peak transitions) or that the WD had turned off.  

Monitoring observations of V407 Cyg at centimeter wavelengths with the EVLA have also shown the appearance, after Day 160, of a resolved nonthermal component at a few times $10^{15}$ cm, depending on the distance adopted for the Mira.\footnote{The EVLA spectra are available at the URL:\\   https://safe.nrao.edu/evla/nova/\#v407cyg.}   This is compatible with the location of the gas that we suggest was ejected during the 1930s event.   The shock, if expanding as inferred in Paper I with a simple power law $v \sim t^{-1/3}$, should have reached a distance of a few times $10^{15}$ cm in about one year.  The 2011 Apr. 23 Ond\v{r}ejov spectrum shows that the [N II] lines had developed extended redward emission with a maximum velocity of about +300 km s$^{-1}$, which is the velocity we expect from the shock at that time.\footnote{As an aside we note that from the [O III] and [N II] diagnostics, the density at this velocity is compatible with a compression ratio of about an order of magnitude, indicating that the shock is likely now isothermal, and for this compression ratio we would expect a steep spectrum with an exponent of about -3.3 (Blandford \& Ostriker 1978) and a strong shock of order $10^3$ in pressure ratio. }

  %  correction for LSR to Hel = 12.5 km/s
\section{Summary}
 
 The main features of the phenomena of the 2010 outburst of V407 Cyg were as follows:
 \begin{itemize}
 \item Absorption lines of low ionization species in the Mira wind disappeared by day 70.  This included the disappearance of P Cyg profiles on the narrow emission lines with expansion velocities of about 50 km s$^{-1}$.    The Balmer line absorption, due to the wind seen against the expanding shock, also systematically decreased in strength and had turned into emission by about day 100.  There was a Balmer progression observed in the absorption  component radial velocity observed with the maximum expansion velocity systematically decreasing.
 \item First appearance of narrow [N II] and [O III] nebular emission lines from the extended low density environment was within the first 7 days.  These lines increased in strength, with evidence for recombination by day 70.
 \item Narrow persistent chromospheric emission was present through at least day 150 on many low ionization profiles, especially [O I] and [Ca II].
 \item The shock appears to have reached the outer wind, based on the extended redward wing on the nebula transitions, by about day  90.
 \item The approaching portion of the shock had stalled by about day 80, thereafter remaining at nearly constant radial velocity, about -200 km s$^{-1}$.  There may have been a contribution from a wake formed behind the Mira.  The receding portion continued to decrease in velocity, reaching +300 km s$^{-1}$ by about day 529  and also appearing as an extended wing on the nebular lines. 
 \item The [Fe X] line intensity due to the shock tracked the {\it Swift} X-ray emission and should be a useful proxy for the high energy emission in symbiotic systems.
 \item The [S III] 6312\AA\ line tracked the 30 GHz radio continuum emission and suggests that the early radio emission was mainly thermal.  The nonthermal radio component developed later, consistent with the shock reaching material from the long duration event of the 1930s.
 \item The system orientation had the Mira near inferior conjunction with the chromospheric features coming from the shadowed side of the star (relative to the white dwarf and shock).  
 \end{itemize}
The spectroscopic behavior displayed in the V407 Cyg outburst is in many ways similar to fluorescent pumping observed in the $\gamma$-ray burst  GRB 060418 by Vreeswijk et al. (2007).  We have no access to the UV for V407 Cyg but the excited state transitions for which we see strong pumping and absorption are the same as those detected from the circumburst gas in GRB 060418.  Finally, we note that recent high resolution spectroscopic studies of SN Ia by Sternberg et al. (2011) have confirmed the presence of low velocity blueshifted absorption features in the  Na I D lines toward a sample of 35 SN Ia that they interpret as a possible contribution of the wind of the precursor system.   These have velocities (of order 150 km s$^{-1}$) that are similar to those observed in V407 Cyg that we attributed to the circumstellar medium from previous wind-like episodes.  The connection between SyRNe and SN Ia has been recently discussed theoretically and quantitatively by Hernanz \& Jos\'e (2008). Walder et al. (2008, 2010), and Patat et al. (2011), among others, and it is possible that the 2010 outburst of V407 Cyg furnishes further evidence for such a causal link.

    \begin{acknowledgements}
    
SNS acknowledges support from the PhD School ``Galileo Galilei'', Univ. of Pisa and the organizers of the 2011 Asiago Symbiotics meeting for their invitation.  GMW acknowledges support from NASA grant NNG06GJ29G.   PK was supported by ESA PECS grant No 98058.  We warmly thank L. Chomiuk, J. Jos\`e,  J. Mikolajewska, K. Mukai, C. Rossi, J. Sokoloski, and S. Starrfield  for discussions and S. Frimann for his kind help with FIES data reductions.     Some spectra at Ond\v{r}ejov were taken by L. Kotkov\'a , P. \v{S}koda, and J. Polster.  We have made extensive use of the Astrophysics Data System (ADS), SIMBAD (CDS), and the MAST archive (STScI) in the course of this work.  We also thank C. Buil for generously  making his observations of the early outburst publicly available.  
 
\end{acknowledgements}

\end{document}